\documentclass[final,5p,times,twocolumn,11pt]{article}

\usepackage[english]{babel}
\usepackage[utf8]{inputenc}
\usepackage{libertinus}

\usepackage{sfmath}

\usepackage{sectsty}
\allsectionsfont{\centering}

\usepackage{mathbbol}
\usepackage{indentfirst}
\usepackage{microtype}
\usepackage{etoolbox}
\apptocmd{\sloppy}{\hbadness 10000\relax}{}{}
\usepackage{doi}
\usepackage[backend=bibtex, citestyle=numeric-comp, style=chem-angew, articletitle=false, doi=true, url=true]{biblatex}
\usepackage[style=british]{csquotes}
\usepackage{siunitx}
\DeclareSIUnit\C{C}
\DeclareSIPrefix\micro{\text{\textmu}}{-3}
\usepackage{tabularx}
\usepackage{colortbl}
\usepackage{multirow}
\usepackage{supertabular}
\usepackage{colortbl}

\makeatletter
\newcolumntype{g}{!{\color{lightgray}\vrule\@width\arrayrulewidth}}
\makeatother

\usepackage{amsmath}
\DeclareMathOperator*{\argmin}{\mathop{\textrm argmin}}
\DeclareMathOperator*{\argmax}{\mathop{\textrm argmax}}
\usepackage{todonotes}
\usepackage{tcolorbox}

\usepackage{prettyref}
\newrefformat{sec}{Section \nameref{#1}}
\newrefformat{subsec}{Subsection \nameref{#1}}
\newrefformat{eq}{Eq.~\eqref{#1}}
\newrefformat{fig}{Figure \ref{#1}}
\newrefformat{tab}{Table \ref{#1}}

\makeatletter
\renewcommand{\@seccntformat}[1]{}
\makeatother

\usepackage[symbol]{footmisc}

\usepackage[left=1.5cm, right=1.5cm, top=1.785cm, bottom=2.0cm]{geometry}
\usepackage{cuted}

\usepackage{fancyhdr}

\usepackage{lastpage}

\begin{document}

\pagestyle{fancy}
\thispagestyle{plain}
\fancypagestyle{plain}{\renewcommand{\headrulewidth}{0pt}}

\fancyfoot{}
\fancyfoot[RO]{\footnotesize{\sffamily{1--\pageref{LastPage} ~\textbar  \hspace{2pt}\thepage}}}
\fancyfoot[LE]{\footnotesize{\sffamily{\thepage~\textbar\hspace{3.45cm} 1--\pageref{LastPage}}}}
\fancyhead{}
\renewcommand{\headrulewidth}{0pt} 
\renewcommand{\footrulewidth}{0pt}

\begin{strip}
\begin{tabular}{m{0.2cm}p{16.8cm}}

 & \noindent\LARGE{
 \textbf{A Primer on Bayesian Parameter Estimation and Model Selection for Battery Simulators}} \\

\vspace{0.3cm} & \vspace{0.3cm} \\

 & \noindent\large{Yannick Kuhn,\textit{$^{1}$} Masaki Adachi,\textit{$^{2,3}$} Micha Philipp,\textit{$^{1}$} David A. Howey,\textit{$^{2}$} Birger Horstmann\textit{$^{1,z}$} \footnote[0]{\textit{$^{z}$}~E-mail: \href{mailto:birger.horstmann@dlr.de}{birger.horstmann@dlr.de}}} \\

\vspace{0.3cm} & \vspace{0.3cm} \\

 & \textit{$^{1}$~German Aerospace Center, Institute of Engineering Thermodynamics, Wilhelm-Runge-Stra{\ss}e 10, 89081 Ulm, Germany.} \\
 & \textit{$^{2}$~University of Oxford, Department of Engineering Science, Parks Road, OX1 3PJ Oxford, United Kingdom.} \\
 & \textit{$^{3}$~Toyota Motor Corporation, Lattice Lab, Nihonbashi, 103-0022 Tokyo, Japan.} \\

\vspace{1cm} & \vspace{1cm} \\

 & \noindent\small{Physics-based battery modelling has emerged to accelerate battery materials discovery and performance assessment. Its success, however, is still hindered by difficulties in aligning models to experimental data. Bayesian approaches are a valuable tool to overcome these challenges, since they enable prior assumptions and observations to be combined in a principled manner that improves numerical conditioning. Here we introduce two new algorithms to the battery community, SOBER and BASQ, that greatly speed up Bayesian inference for parameterisation and model comparison. We showcase how Bayesian model selection allows us to tackle data observability, model identifiability, and data-informed model development together. We propose this approach for the search for battery models of novel materials.} \\

\end{tabular}
\end{strip}


\section{Introduction}

The electrification of industries with diverse energy demands, such as heavy-duty vehicles, grid storage, aviation, or land and sea freight, requires a diversity of battery chemistries. Li-ion batteries alone cannot serve all applications due to the limited supply of lithium, and may also be unnecessarily costly. %
%
The development of new battery materials and electrode compositions currently suffers from long design feedback loops because the assessment of a new material necessitates the study of its longevity and performance in practice. Several physics-based modelling approaches have been developed to complement long-term experimental confirmation with fast theoretical prediction. A significant challenge lies in aligning models with experimental data.

Bayesian statistics is a successful tool for model parameterisation in the battery community \cite{Sethurajan2019,Aitio2020,Escalante2021,Kuhn2022,Adachi2023a,Xia2023,Stier2024,Vogler2024,Sanin2025,Philipp2025}. The benefit of a Bayesian approach is the ability to work with limited and noisy data, but there are two common issues hindering its application. First, we must consider the slow convergence of popular inference methods such as Markov-Chain Monte Carlo (MCMC) \cite{Aitio2020}. For research questions that rely on complex physics-based models to disentangle measurement signals, MCMC would take significant computation time. Thankfully, several alternative methods have been developed recently and applied by the battery community. For example, since the introduction of EP-BOLFI \cite{Kuhn2022} in 2022, we know that Bayesian techniques exist that can answer these research questions in a much more reasonable timeframe. EP-BOLFI is a sample-efficient method that is significantly faster than MCMC.

Another issue that classical optimisers and established Bayesian optimisation approaches have is the lack of a self-assessment criterion. This leads to parameterisations that appear to be precise even when the model is not appropriate for the data, which is a well-known issue in the battery community \cite{Landesfeind2019, Escalante2021}. To address this issue, one must turn to model comparison techniques, but these can be computationally prohibitive and challenging (or sometimes impossible) to implement with MCMC or EP-BOLFI \cite{Philipp2025}.

To enable insight into model selection, we introduce the Bayesian optimisation algorithm SOBER \cite{Adachi2024}, complemented by the Bayesian quadrature algorithm BASQ \cite{Adachi2022}. With these, we leverage uncertainty quantification not only on the scale of an individual parameterisation exercise (given a model and dataset), but also for model selection. This allows us to pursue data-informed model development by assessing the appropriateness of a model given data.

We present a suite of tools built on SOBER to efficiently solve various tasks required for battery research with advanced Bayesian analysis. These include, but are not limited to, selection of a model that best describes given experimental data, actively learned surrogate models that substitute `Big Data' approaches \cite{Aitio2021} with synthetic model evaluations \cite{Dubarry2021}, parameterisation method analysis, and optimisation and parameterisation with uncertainty quantification. To our knowledge, this is the first such toolbox that combines various algorithms in an accessible and suitable manner to enable data-informed model development in battery science. We integrate our front-end to these algorithms into PyBOP~\cite{Planden2024}, a library aiming to harmonise the usage of the various parameter optimisers for the battery community.

We showcase the various benefits of SOBER and BASQ in this paper. First, we give a succinct overview over the theoretical background behind SOBER and BASQ. Then, we present a selection of relevant use cases from current research in batteries. Next, we showcase how to use Bayesian approaches for battery research. Finally, we summarise thoughts on the future use of Bayesian model analysis in battery research.

\section{Background}

\subsection{Understanding parameter fitting as Bayesian inference} 
This introductory section provides a brief overview of Bayesian methods for readers unfamiliar with this approach. Readers already familiar with Bayesian inference may skip this section.

\paragraph{Classical parameter fitting}
We begin with a typical model parameter fitting procedure as an introduction to the Bayesian approach.
First, we establish our notation. Consider a (battery) model \(\mathcal{M}\) simulating some property \(y\) such as voltage response, impedance, or capacity. We denote the input domain of the simulator as \(x\), e.g., frequencies, or measurement timepoints. Observed data are collected as \(\mathbf{D}=\{\mathbf{X}=(x_1,...,x_N), \mathbf{Y}=(y_1,...,y_N)\}\). Finally, we denote the parameters of the simulator as \(\theta=(\vartheta_1,...,\vartheta_d)\), e.g., material properties, or measurement conditions.

Parameter fitting is the task of finding an optimal parameter set \(\hat\theta\) with which the model \(\mathcal{M}\) can accurately reproduce the observed data \(\textbf{D}\). The most common metric to optimise for is the root-mean-squared error (RMSE) between observed and simulated outputs:
\begin{align}
\|\mathcal{M}(\theta, \mathbf{X}) - \mathbf{Y}\|_2 = \sqrt{\frac{1}{N}\sum_{i=1}^N ( \mathcal{M}(\theta, x_i) - y_i )^2}. \label{eq:mse_norm}
\end{align}
We denote the solution of this as:
\begin{align}
    \hat\theta_\text{RMSE} = \argmin_{\theta} \|\mathcal{M}(\theta, \mathbf{X}) - \mathbf{Y}\|_2. \label{eq:mse}
\end{align}

\paragraph{Maximum likelihood parameter fitting}
The preceding approach aligns closely with standard practice, but why is it justified? The key lies in maximum likelihood estimation (MLE), 
a standard tool for parameter estimation in frequentist statistics. Under certain assumptions, the parameter estimated by minimising RMSE (Eq.~\ref{eq:mse}) is equivalent to that obtained via MLE. Therefore, RMSE-based estimation is valid if the assumptions underpinning MLE hold.

The next question is: what are these assumptions, and how does MLE lead to RMSE? The first assumption is the existence of a unique true parameter $\theta^\star$, stable against slight variations of the data, which is a fundamental premise of the frequentist approach. The second is that the observation noise is independent and identically distributed (i.i.d.) Gaussian noise, such that:
\begin{align}
    y_i = \mathcal{M}(\theta^\star, x_i) + \epsilon_i, \quad \epsilon_i\sim\mathcal{N}(0,\sigma^2).
\end{align}
As $\mathcal{M}(\theta^\star, x_i)$ is deterministic and constant, the conditional distribution of individual observations becomes:
\begin{align}
    y_i \sim \mathcal{N}(\mathcal{M}(\theta, x_i), \sigma^2).
\end{align}
From this we define the \emph{likelihood}—the probability of observing output $y_i$ given input $x_i$ and parameters $\theta$:
\begin{align}
    \mathcal{L}(y_i \mid \theta, x_i) = \frac{1}{\sqrt{2 \pi \sigma^2}} \exp \left( - \frac{(\mathcal{M}(\theta, x_i) - y_i)^2}{2 \sigma^2} \right),
\end{align}
This can also be denoted $\mathcal{L}(D_i \mid \theta)$. For the joint likelihood of an entire dataset $\textbf{D} = \{D_i\}_{i=1}^N$, the independent probabilities multiply (since each is i.i.d.):
\begin{align}
    \mathcal{L}(\textbf{D} \mid \theta) = \prod_{i=1}^N \mathcal{L}(y_i \mid \theta, x_i).
\end{align}

Since a higher-valued joint likelihood\footnote{While some literature refers to this joint probability as \textquote{the likelihood}, we also use the term to describe the individual observation likelihoods $p(y_i \mid \theta, x_i)$.} corresponds to a better explanation of the observed data, maximising the joint likelihood yields the most plausible estimate of parameters $\theta$ that could have generated dataset $\textbf{D}$:
\begin{align}
    \hat{\theta}_\text{MLE} = \argmax_{\theta, \sigma}  \prod_{i=1}^N \mathcal{L}(D_i \mid \theta). \label{eq:mle}
\end{align}

This procedure is known as maximum likelihood estimation (MLE). However, working directly with the product of probabilities is numerically unstable, due to underflow in machine precision. To address this, it is standard practice to apply a monotonic transformation—namely, the logarithm—that preserves the location of the optimum. Consequently, MLE is typically performed by minimising the negative log-likelihood.
\begin{equation}
\begin{aligned}
\hat\theta_\text{MLE} =& \argmin_{\theta, \sigma} \sum_{i=1}^N \left[ \frac{(\mathcal{M}(\theta, x_i) - y_i)^2}{2 \sigma^2} + \frac{1}{2} \log(2 \pi \sigma^2) \right]. \label{eq:mle_expand}
\end{aligned}
\end{equation}

Voilà—the least-squares term naturally appears in the numerator of the first term in the objective function. Since taking the square root is also a monotonic transformation, it does not affect the location of the minimum. Therefore, MLE yields the same parameter estimate as RMSE, i.e., $\hat{\theta}_\text{RMSE} = \hat{\theta}_\text{MLE}$ (under the aforementioned assumptions). This reveals that minimising RMSE is equivalent to maximising the likelihood of the observed data $\textbf{D}$. In addition, MLE also provides an estimate of the observation noise variance $\sigma^2$. 

\paragraph{Bayesian inference}
Although MLE provides a suitable explanation for RMSE, it still relies on a strong assumption: the existence of a `true' parameter $\theta^\star$, uniquely capable of accurately predicting experimental outcomes \(\mathbf{D}\) with a given battery model \(\mathcal{M}\). However, we acknowledge that any model is a simplified approximation of the many interlinking physical processes happening inside a battery. Therefore, we should treat the model $\mathcal{M}$ as an assumption rather than absolute truth.

The Bayesian approach reflects this assumption by viewing the model \(\mathcal{M}\) probabilistically. Previously, we implicitly treated the model as a deterministic function, mapping given inputs \(x, \theta\) to outputs \(y\) in a one-to-one manner. A probabilistic perspective, however, suggests that there is no unique true parameter \(\theta^\star\). Instead, we assume \(\theta\) to be a random variable with some underlying probability distribution \(p(\theta)\). The model then maps this to a probability distribution of outputs \(y\).
This perspective may initially seem counterintuitive. One might question why the parameters of a deterministic simulator should be treated probabilistically. The key distinction is that we are addressing an \textit{inverse problem}, mapping observed data $(x,y)$ back to underlying model parameters $\theta$. Such inverse problems often encounter the well-known challenge of identifiability. Non-identifiability refers to the inability to uniquely estimate consistent parameters from given observations under frequentist assumptions---a condition mathematically proven to affect certain battery models \cite{Bizeray2019}. While the nonlinearity of battery models \(\mathcal{M}\) may lead to multiple distinct solutions, non-identifiability refers to poor numerical conditioning when we linearise \(\mathcal{M}(\theta, \mathbf{X})\) to \(\tilde{\mathcal{M}}\) around an optimum:
\begin{equation}
    2 \tilde{\mathcal{M}}(\mathbf{X}) (\tilde{\mathcal{M}}(\mathbf{X}) \hat{\theta}_\text{MLE} - \mathbf{Y}) = \mathbf{0} \Rightarrow \hat{\theta}_\text{MLE} = \tilde{\mathcal{M}}(\mathbf{X})^{-1} \mathbf{Y}.
\end{equation}
Consequently, we interpret these ill-posed conditions as indicating that inferring a unique deterministic parameter $\theta^\star$ is fundamentally impossible. Treating $\theta$ as a random variable and focusing on its statistical characteristics provides a more practical solution for downstream applications. This represents the core rationale behind Bayesian parameter estimation.

Formally, Bayesian inference is based on the following equation, known as Bayes' rule:
\begin{align}
p(\theta \mid \textbf{D}, \mathcal{M}) = \frac{p(\textbf{D} \mid \theta, \mathcal{M})p(\theta)}{p(\textbf{D} \mid \mathcal{M})}, \label{eq:bayes}
\end{align}
where $p(\theta \mid \textbf{D}, \mathcal{M})$ is the \textit{posterior} distribution, that is, the conditional probability distribution of parameters given dataset $\textbf{D}$ and model $\mathcal{M}$. The \textit{likelihood} is commonly denoted by \(p(\textbf{D} \mid \theta, \mathcal{M})\), for which we may again choose \(\mathcal{L}(\mathbf{D} \mid \theta)\), identical to the MLE objective. The term $p(\theta)$ represents the \textit{prior} distribution, reflecting prior trial information or physical constraints about the parameter. The denominator integrates the likelihood over the prior to normalise the right-hand-side to a proper probability distribution summing up to unity:
\begin{equation}
    p(\textbf{D} \mid \mathcal{M}) = \int p(\textbf{D} \mid \theta, \mathcal{M}) p(\theta) \text{d}\theta.
\end{equation}
This denominator is called the \textit{evidence} or \textit{marginal likelihood}; it is the conditional probability that the model generates the observed dataset and is frequently used for model selection. When multiple candidate models $\mathcal{M}_1, \mathcal{M}_2, ...$ exist, the model with the highest evidence should be chosen, as it has the greatest probability of producing the observed dataset---a method known as Bayesian model selection. However, evaluating this integral may be expensive. 

Although the primary goal of Bayesian parameter inference is to estimate the posterior distribution $p(\theta \mid \textbf{D}, \mathcal{M})$ instead of a single parameter $\theta^\star$ estimate, we sometimes desire a Bayesian point estimate. This is achieved by identifying the maximum of the posterior, and can be formulated in terms of the log-posterior. Since the evidence is constant in \(\theta\), this results in:
\begin{align}
\begin{split}
\hat{\theta}_\text{MAP} = \argmax_\theta &\log\left(p(\textbf{D} \mid \theta, \mathcal{M}) p(\theta)\right),\\
= \argmax_{\theta, \sigma} &\sum_{i=1}^N \left[ -\frac{(\mathcal{M}(\theta, x_i) - y_i)^2}{2 \sigma^2} - \frac{\log(2 \pi \sigma^2)}{2} \right] \\
& + \log p(\theta).
\end{split} \label{eq:map}
\end{align}
This approach, known as maximum a posteriori (MAP), adds the log-prior as a regularisation term to the MLE in \prettyref{eq:mle_expand}. This ensures that the resulting $\hat{\theta}_\text{MAP}$ incorporates both our prior assumptions from $p(\theta)$ as well as the impact of new data. Imposing our prior assumptions this way may seem arbitrary at first. But the impact is similar to using regularisation in frequentist regression. In fact, a diagonal and homogeneous Gaussian prior \(\mathcal{N}(0, \tau^2 \mathbb{1})\) with fixed measurement noise variance \(\sigma^2\) reproduces the ridge regression formula with regression factor \(\sigma^2 / \tau^2\):
\begin{align}
\begin{split}
    \hat{\theta}_\text{MAP} = \argmin_{\theta}\Bigg(& \underset{\text{constant w.r.t. } \theta}{\underbrace{\sigma^2 N \log(2 \pi \sigma^2) + \sigma^2 \log\left( 2 \pi \tau^2\right)}} \\
    &+ \sum_{i=1}^N (\mathcal{M}(\theta, x_i) - y_i)^2 + \frac{\sigma^2}{\tau^2} \theta^T \theta \Bigg).
\end{split}
\end{align}
Note that it is usual to ignore the normalisation terms (first two terms) since they are not functions of $\theta$.

\subsection{Approximate Bayesian inference via Bayesian optimisation}

While Bayesian inference provides an appealing solution, solving \prettyref{eq:bayes} is far from straightforward in many models. Conventional methods, such as minimising RMSE outlined in \prettyref{eq:mse}, and MLE in \prettyref{eq:mle}, can be delivered using common optimisers like stochastic gradient descent provided a differentiable simulator is available. In contrast, Bayesian inference is not inherently an optimisation task; rather, it resembles constructing a histogram for statistical analysis by drawing samples. Another way to put this is that conventional approaches are concerned with finding the `best' single-point estimate, whereas Bayesian approaches aim to average (or integrate) over all possible estimates. A widely used technique for this integration is Markov-Chain Monte Carlo (MCMC~\cite{geyer1992practical}).

Although MCMC comes with strong theoretical guarantees, it typically requires numerous samples. In the context of parameter fitting, this translates into numerous model simulations with varying parameter values $\theta$. Physics-based simulators, such as 
DFN models, are computationally expensive to run, rendering MCMC often impractical due to excessive computational cost unless models can be massively simplified and sped up \cite{Aitio2020}. Although we can use MCMC to infer parameter posterior distributions on arbitrary models by brute force, ideally, we would like to perform Bayesian inference on complex models as effortlessly and quickly as standard optimisation tasks using modern techniques.

\paragraph{Likelihood-free inference}
A major challenge in our setting arises from the intractability of the likelihood function for expensive models. While we denote the simulator as $\mathcal{M}$ for convenience, this abstraction hides the true nature of battery simulators, which typically approximate the solutions to partial differential-algebraic equations using numerical techniques such as finite element or spectral element methods. In such cases, $\mathcal{M}$ only yields approximate outputs, and no closed-form expression for the likelihood is available. 

\begin{figure}
    \centering
    \includegraphics[width=\columnwidth]{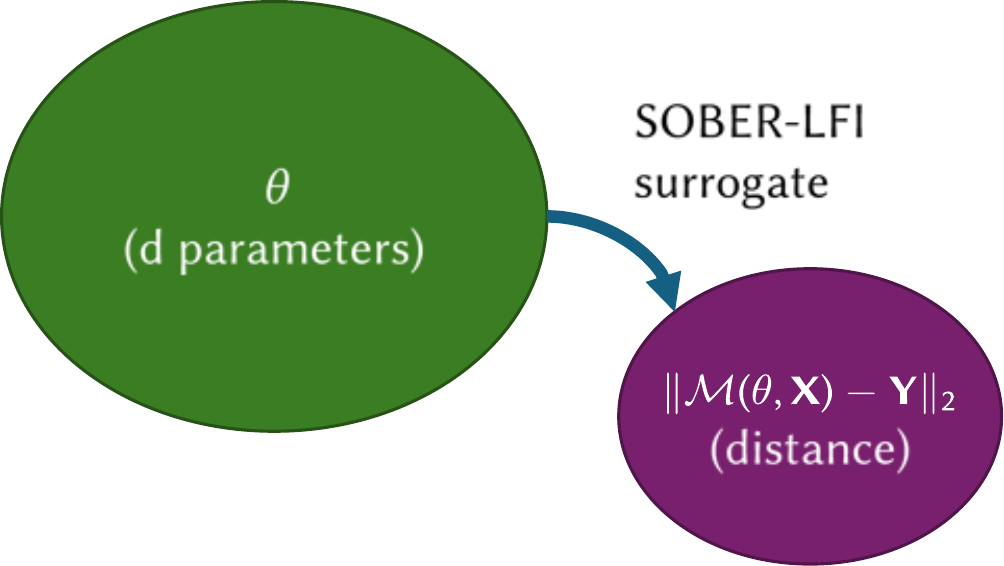}
    \caption{Illustration of the relation between model and LFI likelihood surrogate, which here is calculated via SOBER.}
    \label{fig:surrogate_lfi}
\end{figure}

Likelihood-free inference (LFI)---also known as approximate Bayesian computation (ABC) \cite{diggle1984monte, rubin1984bayesianly, marin2012approximate} or indirect inference \cite{genton2003robust}---is a principled framework designed to address this intractability. Rather than relying on an exact likelihood, LFI introduces a `pseudo-likelihood': an alternative expression that is tractable yet provably converges to the true likelihood as its approximation becomes more precise. As shown in \prettyref{fig:surrogate_lfi}, we may then use this learned pseudo-likelihood as a surrogate for the true relation between model parameters and model fit.

Let us define the discrepancy as $\Delta(\theta) := \sqrt{\frac{1}{N} \sum_{i=1}^N (\mathcal{M}(\theta, x_i) - y_i)^2}$. 
This is the RMSE, and shares the same minimiser, i.e., $\hat{\theta}_\text{RMSE} = \argmin_\theta \Delta(\theta)$. 
Using this, we define the pseudo-likelihood as:
\begin{equation}
    \mathcal{L}_\text{LFI}(\theta) := p(\Delta(\theta) \leq \epsilon) = \int_{-\infty}^\epsilon p (\Delta \mid \theta) \text{d} \Delta
    , \label{eq:lfi}
\end{equation}
where $\epsilon$ is a user-defined threshold.

You may wonder how this definition connects back to the true likelihood. To build intuition: recall from classical parameter estimation that increasing the number of observations usually improves the accuracy of parameter estimates because noise gets averaged out. Bayesian inference follows the same logic---Bayes’ rule updates the prior $p(\theta)$ to the posterior $p(\theta \mid \mathcal{D}, \mathcal{M})$ based on new data via the likelihood. As more observations are incorporated, the posterior \textquote{contracts} around the true parameter and the influence of the prior decreases.
A similar effect occurs in the pseudo-likelihood. A larger $\epsilon$ yields a broader acceptance region, admitting many suboptimal parameters. Conversely, a smaller $\epsilon$ tightens this region, focusing the posterior on more accurate estimates. Thus, adaptively shrinking $\epsilon$ plays a role analogous to accumulating data in classical Bayesian inference. Theoretical results confirm this intuition: under suitable conditions and as $\epsilon \to 0$, the pseudo-likelihood converges to the true likelihood \cite{wilkinson2013approximate, frazier2018asymptotic}.

\paragraph{Gaussian process pseudo-likelihood}
What then is the benefit of the pseudo-likelihood formulation? It becomes particularly powerful when combined with Gaussian process (GP) models \cite{williams2006gaussian}, which provide closed-form expressions for predictive distributions. GPs assume that all marginal and conditional distributions are Gaussian, which simplifies approximation procedures.  Given a discrepancy dataset $\textbf{D}_\Delta := \{(\theta_i, \Delta(\theta_i)),\ i=1,...,T\}$ for $T$ simulation trials, we model $p(\Delta_T \mid \theta) \sim \mathcal{GP}(m_T, C_T \mid \textbf{D}_\Delta) =\mathcal{N}(\Delta(\theta); m_T(\theta), C_T(\theta, \theta))$, where $m_T(\theta)$ and $C_T(\theta, \theta')$ denote the predictive mean and covariance functions. Substituting this into Eq.~\eqref{eq:lfi}, we find:
\begin{align}
    \mathcal{L}_\text{LFI}(\theta) = \Phi \left( \frac{\epsilon - m_T(\theta)}{\sqrt{C_T(\theta, \theta)}} \right), \label{eq:lfi-gp}
\end{align}
where $\Phi$ is the cumulative distribution function (CDF) of the standard normal distribution. This yields a closed-form expression for the pseudo-likelihood, thanks to Gaussianity. 

\paragraph{Updating $\epsilon$ via Bayesian optimisation}
An important remaining question is how to set and update $\epsilon$. Gutmann et al.\ \cite{Gutmann2016} propose a pragmatic strategy: use the minimum discrepancy observed so far, $\epsilon = \min_{i \in [T]} \Delta_{\theta_i}$.
This avoids setting $\epsilon$ too stringently (which would reject all samples) or too loosely (which would accept implausible parameters). As new simulations are run and $\textbf{D}_\Delta$ grows, this threshold naturally tightens, sharpening the pseudo-likelihood and improving approximation to the true likelihood.

To efficiently explore the parameter space, Gutmann et al.\ further propose using Bayesian Optimisation \cite{garnett2023bayesian}. At each iteration $t$, the next parameter estimate $\theta_t$ is chosen by minimising an acquisition function:
\begin{equation}
    \theta_t = \argmin_{\theta} \alpha(\theta) := \argmin_\theta m_T(\theta) - \beta \sqrt{C_T(\theta, \theta)}. \label{eq:af}
\end{equation}
where $\alpha(\theta)$ is the upper confidence bound for Gaussian process (GP-UCB) acquisition function \cite{srinivas2010gaussian}\footnote{Although the above expression serves as a lower bound in the context of minimisation, it is commonly referred to as GP-UCB, following the terminology of the original paper, which framed the problem as a maximisation task.}, balancing exploration (via $\sqrt{C_T(\theta, \theta)}$) and exploitation (via $m_T(\theta)$). The hyperparameter $\beta$ controls the exploration-exploitation trade-off, and can be set based on theoretical guarantees \cite{chowdhury2017kernelized}.

As more simulations are run, $\epsilon$ becomes increasingly smaller, the GP model becomes more accurate, and the pseudo-posterior approaches the true posterior. Once convergence is detected (e.g., $\theta_t$ no longer improves), we can obtain both the MAP estimate and the approximate posterior:
\begin{equation}
\begin{aligned}
    \hat{\theta}_\text{MAP} &\approx \argmin_{\theta} m_T(\theta),\\
    p(\theta \mid \textbf{D}, \textbf{D}_\Delta, \mathcal{M}) &\propto \mathcal{L}_\text{LFI}(\theta) p(\theta).
\end{aligned}
\end{equation}
GP-UCB is well-known for its sample efficiency, achieving near-optimal solutions with a remarkably small number of simulations. Since simulation is typically the most expensive part of Bayesian inference in this setting, minimising the number of runs directly translates into faster and more scalable inference.

\paragraph{Parallelisation of Bayesian optimisation}
Having established that Bayesian inference tasks can be effectively addressed through Bayesian optimisation, a remaining challenge pertains to sequential computation. As illustrated in \prettyref{eq:af}, Bayesian optimisation traditionally selects each query point $\theta_t$ sequentially, necessitating the completion of simulator runs one-by-one. Given that physics-based simulators typically represent the primary computational bottleneck in terms of parallelisation, sequential execution is highly inefficient. To circumvent this, \textquote{batch} Bayesian optimisation has been developed, which selects multiple query points simultaneously. These query points can then be evaluated concurrently using computational resources such as computer clusters, multi-core CPUs, or GPUs. In theory, this provides a linear speedup with respect to batch size. For example, if 100 samples are required to converge to $\theta_\text{MAP}$, and each simulation takes 15 minutes, sequential Bayesian optimisation would require at least a day. With parallel computation on 48 CPU cores, batch Bayesian optimisation would finish within an hour.

However, achieving effective parallelisation is challenging because the next query is traditionally determined by the singular maximiser of an acquisition function. To generalise to batches, we recall that a core rationale of the UCB approach is monotonic reduction of the confidence interval of the optimum. We can re-interpret this goal as aiming for maximal information gain by selecting subsequent points where the GP model variance is high. The selection of points with maximal predictive variance is known as \emph{uncertainty sampling}. Indeed, an acquisition function variant based on uncertainty sampling has proven highly effective for exploratory sampling. This uncertainty-driven approach naturally extends to multiple-point selection, linking it to batch uncertainty sampling and Bayesian quadrature \cite{Adachi2022}. The SOBER method leverages this connection to perform batch Bayesian optimisation using Bayesian quadrature techniques. Further technical details are available in the referenced literature \cite{Adachi2024}.

\subsection{Usage of Bayesian optimisation in practice}

All examples in this paper solve optimisation as Bayesian estimation via recombination (SOBER~\cite{Adachi2024}), with Bayesian alternately subsampled quadrature (BASQ~\cite{Adachi2022}) to perform Bayesian model analysis. The minimal setup consists of model, data, parameter search area, and either a fixed number of model evaluations to attempt with SOBER, or a target log-marginal likelihood to aim for. The latter is calculated via BASQ and gives a model selection criterion that also serves as a convergence criterion with excellent accuracy \cite{Adachi2023a}. If we want to use this criterion, we need one additional number, and that is the number of integration nodes for BASQ to construct. In summary, we only need to give two settings; SOBER and BASQ handle the rest. This allows for quick prototyping on new problems, and facilitates familiarisation with the approach. At the same time, fine-tuning the implementation details is possible and transparent at every level, making popular extension techniques seamlessly available, including but not limited to: prior selection, acquisition functions, custom log-likelihoods, gradient tracking, parallel computing on GPUs, or expectation propagation (EP~\cite{Minka2001}); see the Appendix for a brief introduction of the latter. We refer the interested reader to our previous publications \cite{Adachi2024, Kuhn2022} that go into greater detail about these aspects.

\section{Example applications and results}

We will showcase various examples of currently relevant research topics in battery material science to demonstrate the benefit and flexibility of our methods. To build understanding step-by-step, we introduce one Bayesian model analysis concept at a time. The first four examples are intentionally structurally simple: lifetime optimisation for posteriors, voltage relaxation parameterisation for correlations, knee points for model selection, and model-based data augmentation for inverse modelling. Our method scales well to more complex models, as verified in previous publications \cite{Adachi2024, Kuhn2022}. To reiterate this claim, we also investigate electrolyte identifiability through rate tests, and selection of physics-based models from impedance data.

\subsection{Example 1: Battery sizing}

\paragraph{Method}

In this first example, we use a Bayesian approach to find posterior distributions of optimal battery sizing parameters for a solar-battery system, considering the trade-off between battery lifetime and energy costs. %
%
%
Battery lifetime depends on many factors \cite{Reniers2019}. Note that we define lifetime here as the days until the battery degrades to 80\% of initial capacity. If we assume reasonable mechanical and electrical conditions, the most influential factors are calendar age and usage pattern. To model these effects in a succinct, yet realistic fashion, we only consider the build-up of the solid electrolyte interphase (SEI) \cite{vonKolzenberg2020}, which consumes the usable battery capacity.

What can we infer from modelling the build-up of SEI? Imagine you have a solar panel and want to couple a battery to it to provide electricity at nighttime. We'll assume the overnight power usage is constant and that the solar panel is situated near the equator with steady weather patterns, so that we may model each day with the same average battery cycling pattern. A battery that has exactly the nightly required capacity will cycle between 0\% and 100\% state of charge (SOC) daily.

However, we know that the SEI grows much faster close to 100\% SOC. So one might prefer a slightly larger than necessary battery to only cycle, e.g., between 0\% and 90\% SOC. Up to a certain oversize factor, the extra lifetime will be greater than the extra cost the extra capacity entails. So we wish to find the optimal cost-benefit battery size.
%
As a search area for the optimal size we take between 100\%  and 120\% of the minimum required battery capacity. To formulate this for Bayesian methods, we take a uniform prior parameter distribution over the interval [100\%, 120\%]. We take 16 initial model samples and then run 7 SOBER iterations with 16 samples each. The BASQ evaluation of the posterior is set up with 128 integration nodes. The battery parameters are from the example battery in the legacy DUALFOIL code \cite{Newman2007} of a graphite-cobalt oxide cell, to which we added the SEI growth parameters from von Kolzenberg et al.\ \cite{vonKolzenberg2020}.

\paragraph{Result and Discussion}

\begin{figure}
    \centering
    \includegraphics[width=\columnwidth]{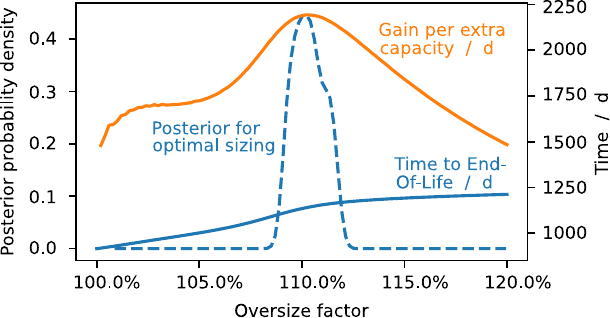}
    \caption{Extra lifetime of a solar panel battery oversized to avoid high charge states. Time to end-of-life (blue, solid) increases with extra capacity, but there is an optimum lifetime gain per extra capacity (orange) at around 10\% extra. Bayesian optimisation  estimates a posterior (blue, dashed) that encodes the most relevant part of the gain function.}
    \label{fig:optimisation}
\end{figure}

We aim to size a solar panel battery for optimal cost-lifetime trade-off as our first example, with results shown in \prettyref{fig:optimisation}. The battery lifetime (blue line) increases with extra capacity, but we have diminishing returns beyond a certain point. In this one-dimensional example, we can trivially find without any Bayesian tools that the optimal cost-benefit size is at 10\% extra capacity. Note that this requires that a dense sampling of the gain function (orange line) is feasible, which will not be true for higher-dimensional parameterisations. So we introduce the Bayesian posterior (blue dashed line) here, which we can also efficiently compute for higher-dimensional use-cases. The posterior probability density function (PDF) describes a range of probable optima.
%
The posterior is purposely a function that is different from the gain function. While SOBER learns a surrogate for the gain function, for Bayesian optimisation we prune the surrogate to only its most likely part. This shows that the optimal extra capacity is at around 10\%.

\subsection{Example 2: Voltage relaxation}

\paragraph{Method}

Parameter sensitivity can be examined in a Bayesian framework by considering the posterior probability distributions of the fitted parameters. In this second example we introduce the covariance of the posterior as a parameter sensitivity analysis approach. %
We explain correlation matrices and covariance plots, and how to gauge parameter identifiability from them.

Covariances describe the relations between any two parameters in an arbitrarily high-dimensional probability distribution. 
Think of the covariance plot as a more fine-grained and global extension of sensitivity analysis. We may reduce this matrix of plots to a matrix of scalars, the covariance matrix, e.g., by regressing a normal distribution. This covariance matrix contains both parameter accuracies and their interdependencies. To consider parameter identifiability regardless of parameter accuracy, we may express the \textquote{tilt} in each cell of the covariance plot via correlations. Correlations \(C(p_i, p_j)\) are a normalised view of covariances \(\sigma_{i,j}^2\), considering the standard deviations \(\sigma_i\) and \(\sigma_j\) of variables \(i\) and \(j\):
\begin{equation}
    C(p_i, p_j) = \frac{\sigma_{i,j}^2}{\sigma_i \sigma_j}.
\end{equation}

A correlation near zero between two parameters shows that they are uniquely identifiable from each other; correlations near +1 or -1 show that parameter pairs are interchangeable within their errorbars. 
Finally, to visualise the impact of high-dimensional parameterisations, we may also employ the predictive posterior. The predictive posterior is the model \(\mathcal{M}\) evaluated on the parameterisation posterior \(p(\theta \mid \mathbf{D}, \mathcal{M})\): \(\mathcal{M}(p(\theta \mid \mathbf{D}, \mathcal{M}))\). We may also think of this as the model-informed posterior of the data.  

Voltage relaxation in batteries refers to their rather slow equilibration after use. Especially in batteries with silicon-based electrodes, the measured voltage of a battery may slowly change for up to multiple weeks after use \cite{Wycisk2024}. This may be mistaken for reduced remaining energy, or it may disturb charging protocols. As an example, we plot a galvanostatic intermittent titration technique (GITT) \cite{Kang2021} measurement in \prettyref{fig:wycisk} that shows silicon electrode voltage relaxations at multiple states of charge, each for about a day \cite{Wycisk2024}. We will focus on a single of these charge states only. 
To improve battery diagnostics, we model the voltage relaxation with the model from Köbbing et al.\ \cite{Köbbing2024}, which has the following form with a logarithmic slope \(\text{log-slope}\) and exponential relaxation timescale \(\tau\):
\begin{align}
\begin{split}
    \Delta U(t) &= \text{log-slope} \cdot \text{arctanh} \bigg( \\
    &\tanh\left(\frac{\Delta U(t=\infty)}{\text{log-slope}}\right) \cdot  \exp\left(-\frac{t}{\tau}\right) \bigg).
\end{split}
\end{align}

In this model, the voltage relaxation behaves logarithmically with time for a long time, making it virtually impossible to predict the terminal voltage early.

We will analyze the accuracy with which the terminal voltage \(\Delta U(t = \infty)\) of the battery cell can be predicted, and how relevant the log-slope and relaxation timescale parameters are for this prediction. 
We dissect the parameterisation task using expectation propagation (EP) \cite{Minka2001} into two time frames in the relaxation, one between 10 and 10000 seconds, and one from 10000 seconds onward. Note that EP has a natural tendency to automatically prefer features that are most informative \cite{Kuhn2022}. This may help SOBER more easily discard the short timescale data, but only if necessary to fit the long timescale. More importantly, since each timescale by itself has a lower dimensionality, fewer model samples are required for SOBER to find the optimum in our very wide initial search area. For this benefit, wo do incur the risk of missing a global optimum in a multimodal true posterior; we will need to check the predictive posterior for overconvergence, i.e., we need to check that the data is still contained within the predictive posterior predictions.
\begin{table}
\centering
\caption{Prior for silicon voltage relaxation example---a multivariate log-normal distribution defined such that the 95\% confidence region touches these bounds.}
\label{tab:silicon_relaxation}
\begin{center}
\begin{tabular}{c|cc}
\hline
parameter & transform & bounds \\
\hline
\(\Delta U(t=\infty)\) [\si{\volt}] & log & \([0.01, 0.2]\) \\
log-slope [\si{\volt}] & log & \([0.001, 0.2]\) \\
relaxation timescale [\si{\second}] & log & \([10^3, 10^7]\) \\
\hline
\end{tabular}
\end{center}
\end{table}

The prior for the parameters is defined in \prettyref{tab:silicon_relaxation}. We take 128 initial model samples and then run two EP iterations with four SOBER iterations each at 64 samples each. The dampening of EP is set to give final dampening of the likelihood at 0.5. The BASQ posterior evaluation is set up with 32 integration nodes.

\paragraph{Result and Discussion}

\begin{figure}
    \centering
    \includegraphics[width=\columnwidth]{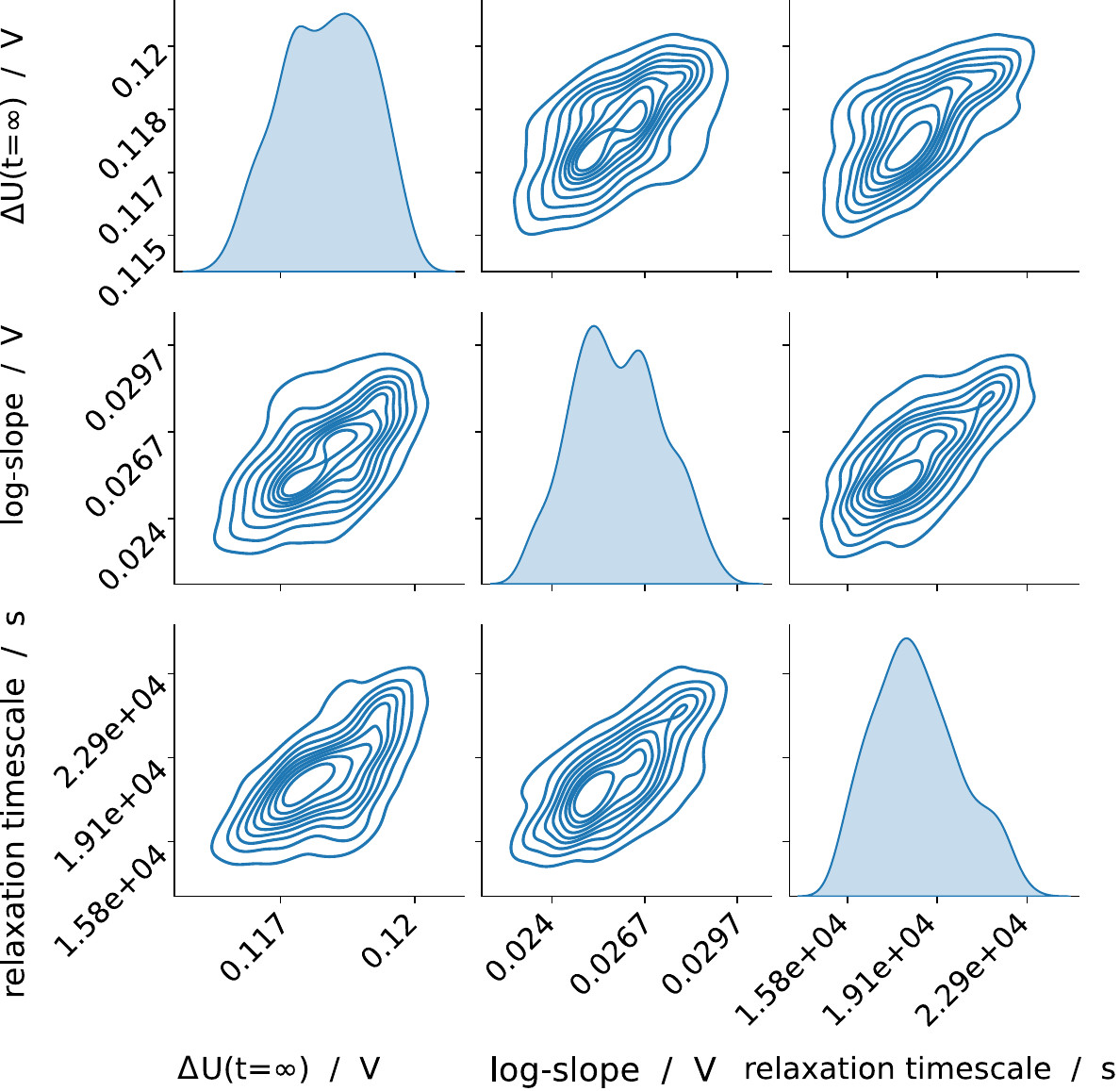}
    \caption{Parameterisation posterior for a silicon electrode voltage relaxation model. We employ a visualisation suitable for several dimensions: the diagonal shows marginal distributions of each individual parameter; off-diagonals show joint distributions of parameter pairs. Joint distributions are visualised as contour plots, with each contour denoting one shared probability value.}
    \label{fig:parameterisation_posterior}
\end{figure}

\begin{figure}
    \centering
    \includegraphics[width=0.8\columnwidth]{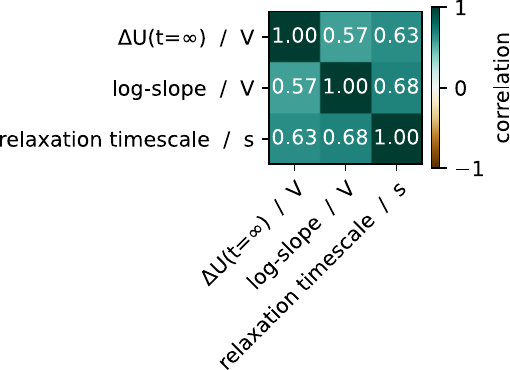}
    \caption{Correlation matrix of  multivariate normal approximation of \prettyref{fig:parameterisation_posterior}. 
    To emphasise parameter interchangeability, we show off-diagonal covariances divided by square-root of  the corresponding variances on the diagonal.
    }
    \label{fig:parameterisation_correlation}
\end{figure}

\begin{figure}
    \centering
    \includegraphics[width=\columnwidth]{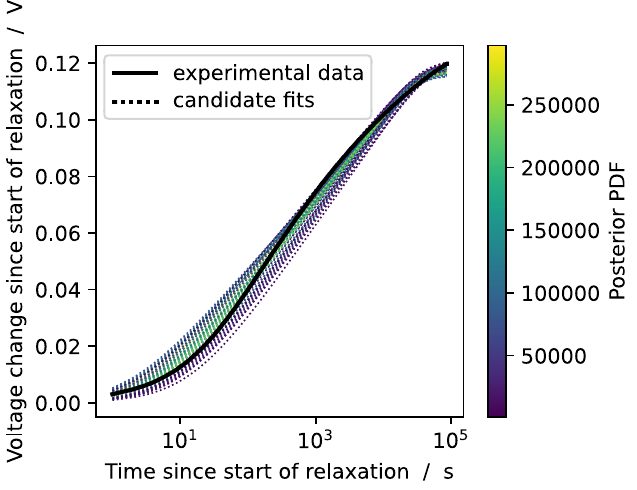}
    \caption{Measurement of silicon electrode voltage relaxation, alongside  representative model evaluations from the parameterised model. These are coloured according to the posterior probability density function.}
    \label{fig:parameterisation_predictive}
\end{figure}

In the covariance plot in \prettyref{fig:parameterisation_posterior}, we see one- and two-dimensional slices through the multi-dimensional posterior parameter distribution. These slices show us that although each parameter by itself has only one optimum, pairs of parameters may vary together. We can reduce the information in this plot to an optimal parameter set (the mean) and the local shape of the posterior around it (the covariance) by regressing a multivariate normal distribution to the posterior distributions. %
To analyze the interchangeability of parameters, we normalise the covariance matrix to the correlation matrix shown in \prettyref{fig:parameterisation_correlation}. Here we see a quantified degree of parameter interchangeability, with values around and above 0.5 indicating high interchangeability within each parameter's error bars. %
The final visualisation is the predictive posterior in \prettyref{fig:parameterisation_predictive}. This shows the various parameterisations that fit the data reasonably well, coloured by how well they do so.

This second example introduces us to the Bayesian approach to sensitivity analysis. The covariance is similar to the inverse of the Hessian in sensitivity analysis and the predictive posterior is similar to the output sensitivity. The covariance plot and correlation matrix help us study the identifiability of the parameters. If there were multiple equally valid local optima, we would see multiple distinct peaks in \prettyref{fig:parameterisation_posterior}, called multimodality. Here, with a unimodal distribution, we can still observe that the parameters are interchangeable to a degree due to their high correlations. The predictive posterior helps us identify the sensitivity and predictive power of the model fit. Since we have interchangeable parameters, we can see how adjusting them together may fit the data better on the short or on the long timescale. We can also confirm that the predictive posterior still contains the experimental data, so we haven't applied EP for longer than appropriate. If we desire tighter convergence, we would take this posterior as the prior for a non-EP run.

\subsection{Example 3: Degradation model selection}

\paragraph{Method}

Evidence in the Bayesian sense is a measure of \textquote{data quality given a model}.\footnote{It is a common mistake to think evidence describes \textquote{how well the model explains the data}. The discussion around \prettyref{eq:evidence} will clarify why that is not completely correct.} In our third application example we convey how to interpret Bayesian evidence in practice. We illustrate its role in Bayesian model selection via knee point identification in battery degradation trajectories. %
%
Knee points are sudden accelerations in the degradation rate over time. Identifying their occurrence and shape early \cite{Zhao2025, Tao2025, Zhu2025} gives crucial information about the remaining useful service life of a battery, especially when considering second-life applications \cite{Attia2022, Ni2025}.

\begin{figure}[!ht]
    \begin{center}
    \includegraphics[width=\columnwidth]{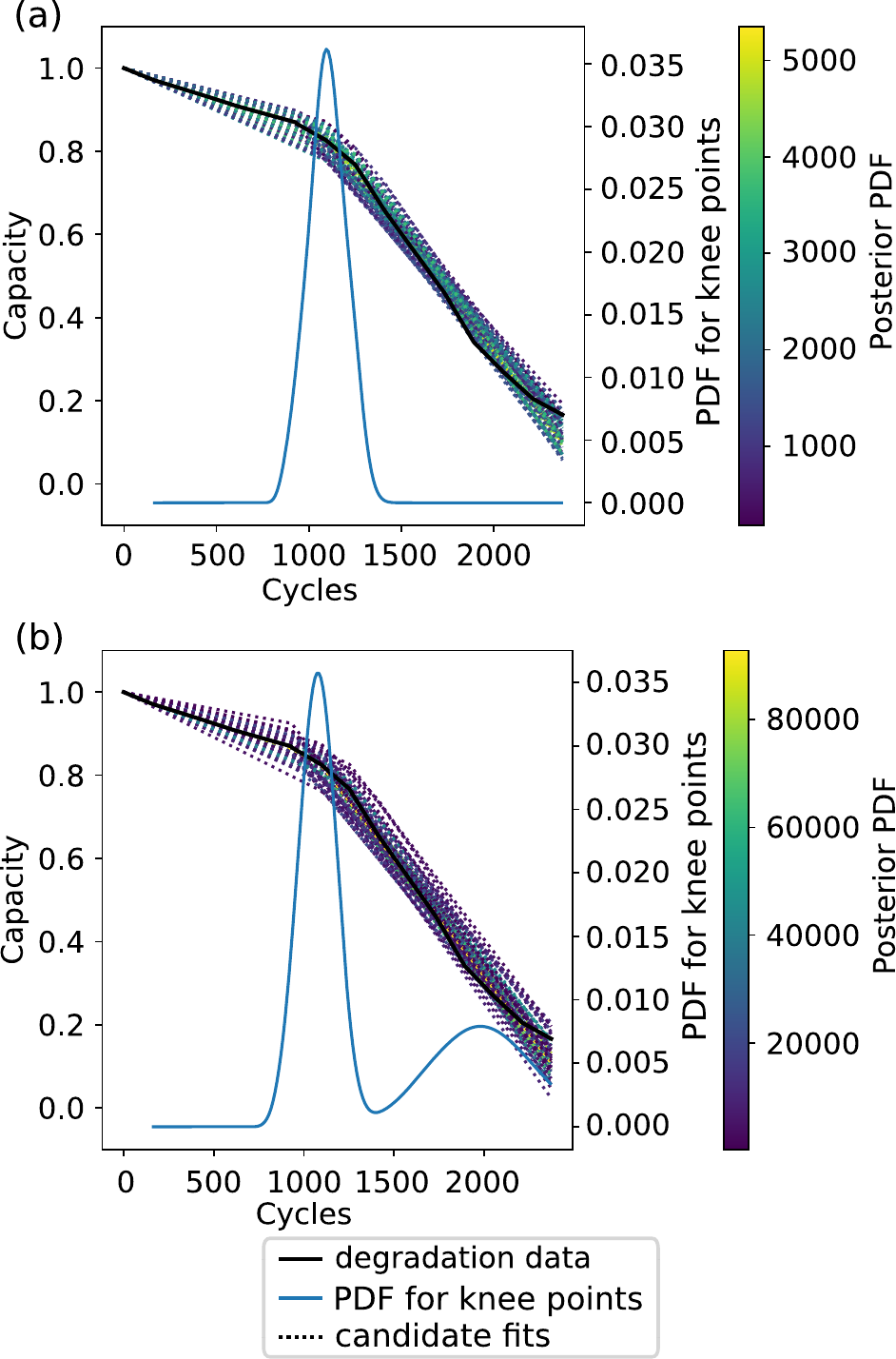}
    \end{center}
    \caption{Model selection example with (a) one-knee point and (b) two-knee point models, in both cases using the same experimental data (black). Model candidate realisations (dashed) are coloured with brighter lines the better they fit. Marginal probability distributions of knee point locations are plotted in blue.}
    \label{fig:model_selection}
\end{figure}

We use battery capacity loss data on 48 identical cells from Baumhöfer et al.\ \cite{Baumhöfer2014}, in the form published by Attia et al.\ \cite{Attia2022}. See the complete dataset in \prettyref{fig:baumhofer}. We model the knee point(s) with a simple piecewise-linear curve with one or two bends. The corresponding parameters are the individual slopes and the points at which they intersect. We choose one representative data curve, \prettyref{fig:model_selection}, and identify the number of knee points by computing and comparing the model evidence for the one-bend vs.\ two-bend models. As Harris et al.~\cite{Harris2017} note, multiple different failure modes may occur over a battery lifetime, motivating more than one knee point.

We set up priors of the locations and in-between slopes for one knee point and two knee point models for comparison, as described in \prettyref{tab:kneepoints}. We took 256 initial model samples and then ran 12 SOBER iterations with 64 samples each. The BASQ evaluation of the posterior was set up with 256 integration nodes.
\begin{table}
\centering
\caption{Prior parameter values for the degradation model selection example. The one knee point model uses only the first three parameters. All priors are multivariate log-normal distributions defined with the 95\% confidence region touching the bounds.}
\label{tab:kneepoints}
\begin{center}
\begin{tabular}{c|cc}
\hline
parameter & transform & bounds \\
\hline
Rel. cap. loss per cycle & \multirow{2}{*}{log} & \multirow{2}{*}{\([10^{-5}, 10^{-3}]\)} \\
before \(1^\text{st}\) knee point & & \\
\arrayrulecolor{gray}\hline
Cycle no. of \(1^\text{st}\) knee point & log & \([200, 2000]\) \\
\hline
Rel. cap. loss per cycle & \multirow{2}{*}{log} & \multirow{2}{*}{\([10^{-5}, 10^{-2}]\)} \\
after \(1^\text{st}\) knee point & & \\
\hline
Cycle no. of \(2^\text{nd}\) knee point & log & \([700, 2500]\) \\
\hline
Rel. cap. loss per cycle & \multirow{2}{*}{log} & \multirow{2}{*}{\([10^{-5}, 10^{-2}]\)} \\
after \(2^\text{nd}\) knee point & & \\
\arrayrulecolor{black}\hline
\end{tabular}
\end{center}
\end{table}

\paragraph{Result and Discussion}

We see the same data parameterised with one or two knee points in \prettyref{fig:model_selection}. There is an evident degradation rate increase when the cell hits 80\% of its initial capacity, but one might argue whether or not there is a degradation rate decrease at 40\%. The model evidences suggest the one-knee point model is the less appropriate model, with evidence \(0.0494 \pm 0.0004\), versus the two-knee point model with evidence \(0.18934 \pm 0.00006\). 

Evidence is the concept we wish to illustrate here---this can quantify how well the model and data fit together when comparing between several models. A higher mean evidence indicates an appropriate model (balancing complexity vs.\ goodness of fit), as broad likelihoods or those with multiple peaks naturally lead to a lower evidence value. The variance of the evidence estimate tells us the accuracy with which BASQ computed the evidence. Assuming we have chosen a sufficient number of integration nodes, this accuracy corresponds to a convergence criterion. To link evidence with the metric we usually value most, the predictive posterior spread, we take a closer look at \prettyref{fig:model_selection}. In the better model in \prettyref{fig:model_selection}(b), any model realisation that does deviate from the data is assigned a \textquote{wrong} / dark colour. In \prettyref{fig:model_selection}(a), no realisation of the one-kneepoint-model can perfectly match the data, so there are several comparably \textquote{right} / bright coloured model realisations that over- or undershoot the data in various places.

To movitate why this quantified evidence is helpful in practice, consider data from a battery that has not yet encountered its first knee point. If we compute the evidence for zero knee points, we would see it gradually decreasing when the battery tips over a knee point. We could even use a predictive model instead of our simplified piecewise linear model, and have it predict the occurrence of a knee point even earlier by observing the increase in evidence shortly before the knee point occurs.

\subsection{Example 4: Inverse modelling via actively learned surrogates}

\paragraph{Method}

\begin{figure}
    \centering
    \includegraphics[width=\columnwidth]{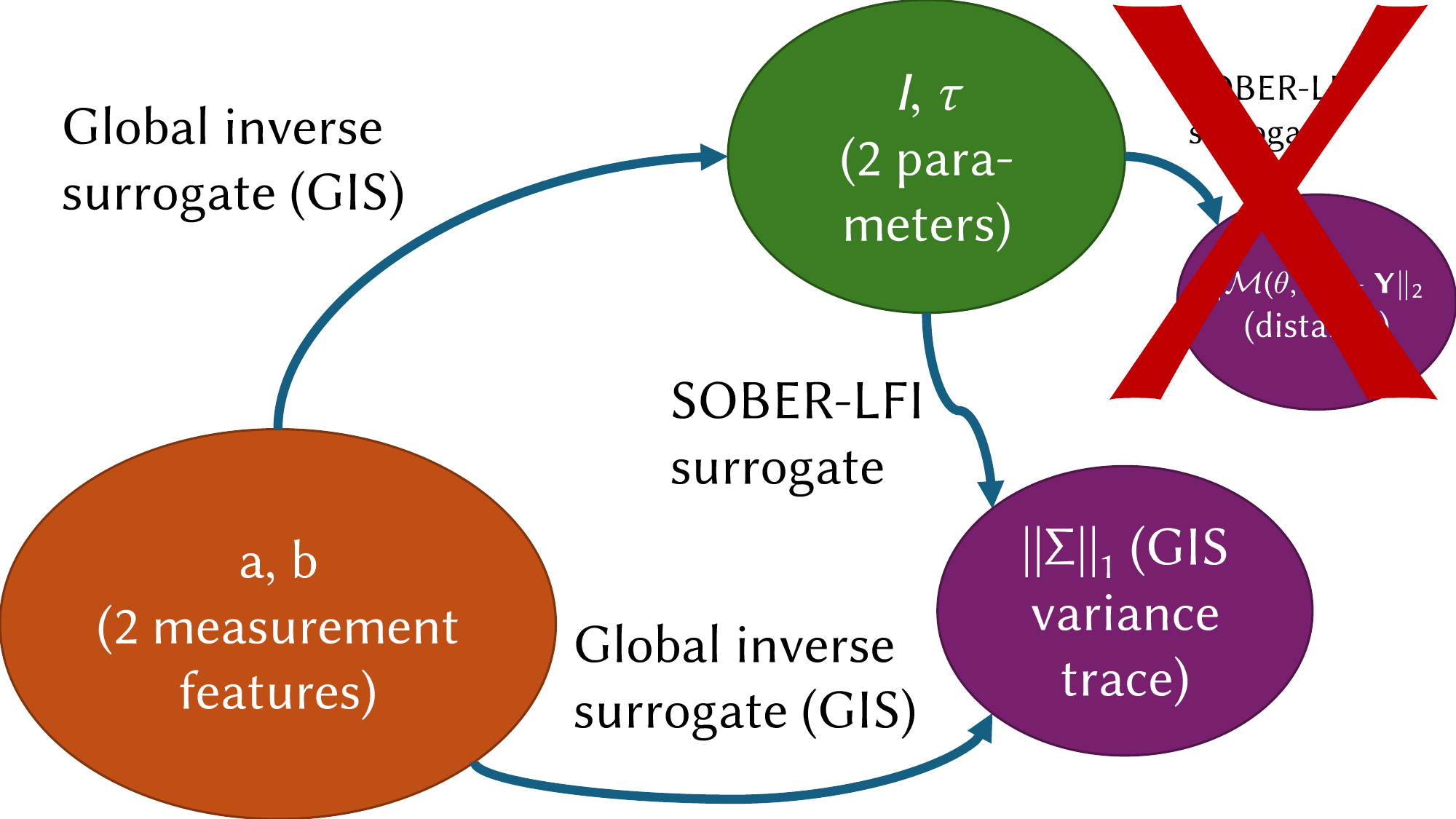}
    \caption{Illustration for the way SOBER is utilised within the global inverse surrogate training loop.}
    \label{fig:surrogate_encapsulation_intro}
\end{figure}

Inverse modelling is also commonly referred to as parameterisation---given a forward model and data that is close to a possible model output, the goal is to find the best model parameters to reproduce the data. We now consider the case of a large set of similar data to be parameterised with a physics-based model. Rather than parameterise each piece of data individually, it may make more sense to train a faster surrogate model to learn the parameterisation task just once. To distinguish this surrogate from the one SOBER trains to approximate the likelihood, we call this  the global inverse surrogate (GIS). The question then becomes, what is the optimal training dataset to minimise expensive model evaluations? In this fourth application example, we leverage SOBER to train a Gaussian process (GP) global inverse surrogate on a toy example: predicting battery operating conditions from incomplete data. The same approach has already been shown to be able to predict battery states from incomplete data \cite{Ko2024}. We show in \prettyref{fig:surrogate_encapsulation_intro} how we implement this approach using SOBER. Compared to \prettyref{fig:surrogate_lfi}, SOBER is now training its LFI surrogate to learn the whole parameterisation task, rather than a single parameterisation.

To reduce the complexity of our toy example, we consider a restricted space of battery operating conditions: constant-current pulses, followed by a rest phase, for a known battery makeup. So the only battery model inputs we vary are the magnitude \(I\) of the applied current and the duration \(\tau\) it is applied for. As the incomplete data output, we consider a short snippet of the voltage relaxation at the beginning of the rest phase. Since we know that it will always be roughly square-root-shaped, we can reduce the output dimension for our parameterisation task to two numbers using least-squares regression: the offset \(a\) of a fitted square-root function, and its slope \(b\):
\begin{equation}
    (I, \tau) \mapsto U(t) \mapsto (a, b) \text{ with } U(t\geq \tau) \approx a + b \sqrt{t}.
\end{equation}

As an additional benefit to this reduction, the global inverse surrogate model will learn how to interpret the shape of the voltage signal, rather than its specific discretisation in time. %
%
Our global inverse surrogate model base of choice is a Gaussian process, which allows us to map from observations to inputs or parameters probabilistically---predicting both mean and covariance, not just point values (e.g., \cite{Wang2022}).

We simulate the short voltage relaxation time series after varying operating conditions with the Doyle-Fuller-Newman model \cite{Doyle1993} using PyBaMM \cite{Sulzer2021}. We take as DFN model parameters the example battery from the legacy DUALFOIL code \cite{Newman2007}.
We then task SOBER to perform an iterative design of experiment feedback loop. Given (initially random) points \((I, \tau)\) and the corresponding outputs \((a, b)\), the global inverse surrogate GP learns the relationship \((a, b) \mapsto (I, \tau)\) from them. SOBER analyses the variance of the GP and designs new points \((I, \tau)\) to globally minimise that variance in the next training step. This assumes, of course, that \((a, b) \mapsto (I, \tau)\) can be modelled via GPs. To be precise, we employ a multitask-multioutput GP with Kronecker structure (i.e., outer product structure) for input \((I, \tau)\) and its fidelity \((f_I, f_\tau)\):
\begin{equation}
    (I, \tau) \sim \mathcal{GP}\left(
    \begin{bmatrix}
        I(a, b) \\ \tau(a, b) \\ f_I(a, b) \\ f_\tau(a,b)
    \end{bmatrix},
    \begin{matrix} K((a, b), (a', b')) \\ \otimes K_f((a, b), (a', b')) \end{matrix}
    \right).
\end{equation}

Through this approach, we expect not only to obtain completed data quickly, but also to receive an accuracy assessment of that completion. %
%
We set up the prior over which to train the GP global inverse surrogate as a uniform prior over constant-current rates in battery capacity per hour between \([0.02, 1.0]\) and durations in seconds between \([1, 600]\). The global inverse surrogate training dataset was generated from 128 initial samples and 3 SOBER iterations at 128 samples each. The acquisition criterion for the SOBER samples is maximum uncertainty, i.e., the maximum uncertainty from the intermediate-trained GP.

\paragraph{Result and Discussion}

\begin{figure}
    \centering
    \includegraphics[width=\columnwidth]{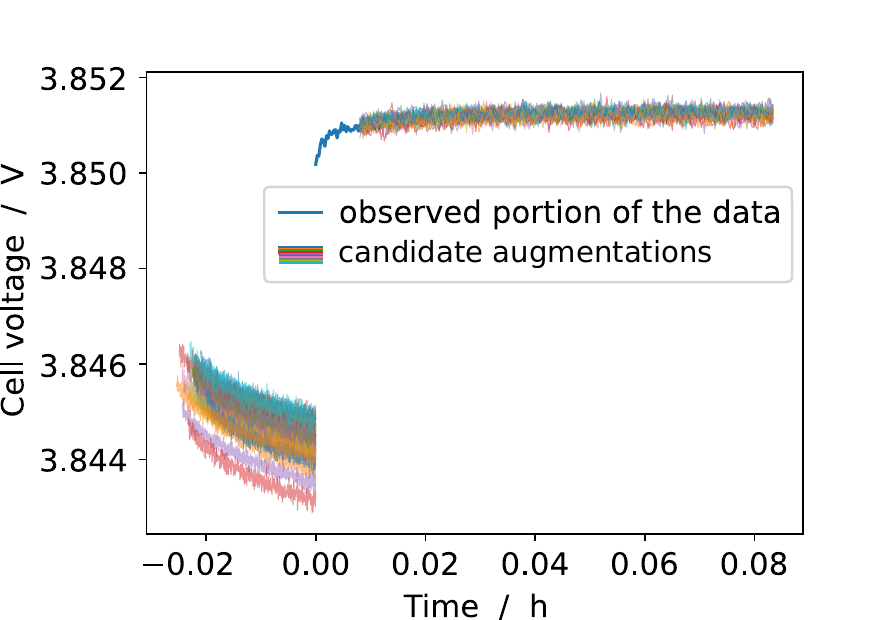}
    \caption{Inverse modelling example with data augmentation. The data (blue) only contains a small fraction of the relaxation phase. The trained inverse model proposes a variety of possible curves (multicoloured) containing the pulse itself and the long-term relaxation.}
    \label{fig:inverse_modelling}
\end{figure}

In \prettyref{fig:inverse_modelling}, we see the predictive posterior (multicoloured, output) of the global inverse surrogate on a snippet of battery voltage relaxation data (blue, input). The posterior itself was generated by taking rapid draws (within a few milliseconds) from the GP posterior and running DFN simulations against each one. We see a considerable range of operating conditions, but they fall within \qty{2}{\milli\volt} of each other, and the relaxation is uniquely identified up to measurement accuracy.

Training a surrogate model for inverse modelling involves exploring the entire prior on a representative sample of data. As Bayesian optimisers tend to thoroughly explore a large region of the prior even for a single parameterisation, we find that training a global inverse surrogate (GIS) does not require significantly more model evaluations.


\subsection{Example 5: Electrolyte transport identifiability}

\paragraph{Method}

\begin{figure}
    \centering
    \includegraphics[width=\columnwidth]{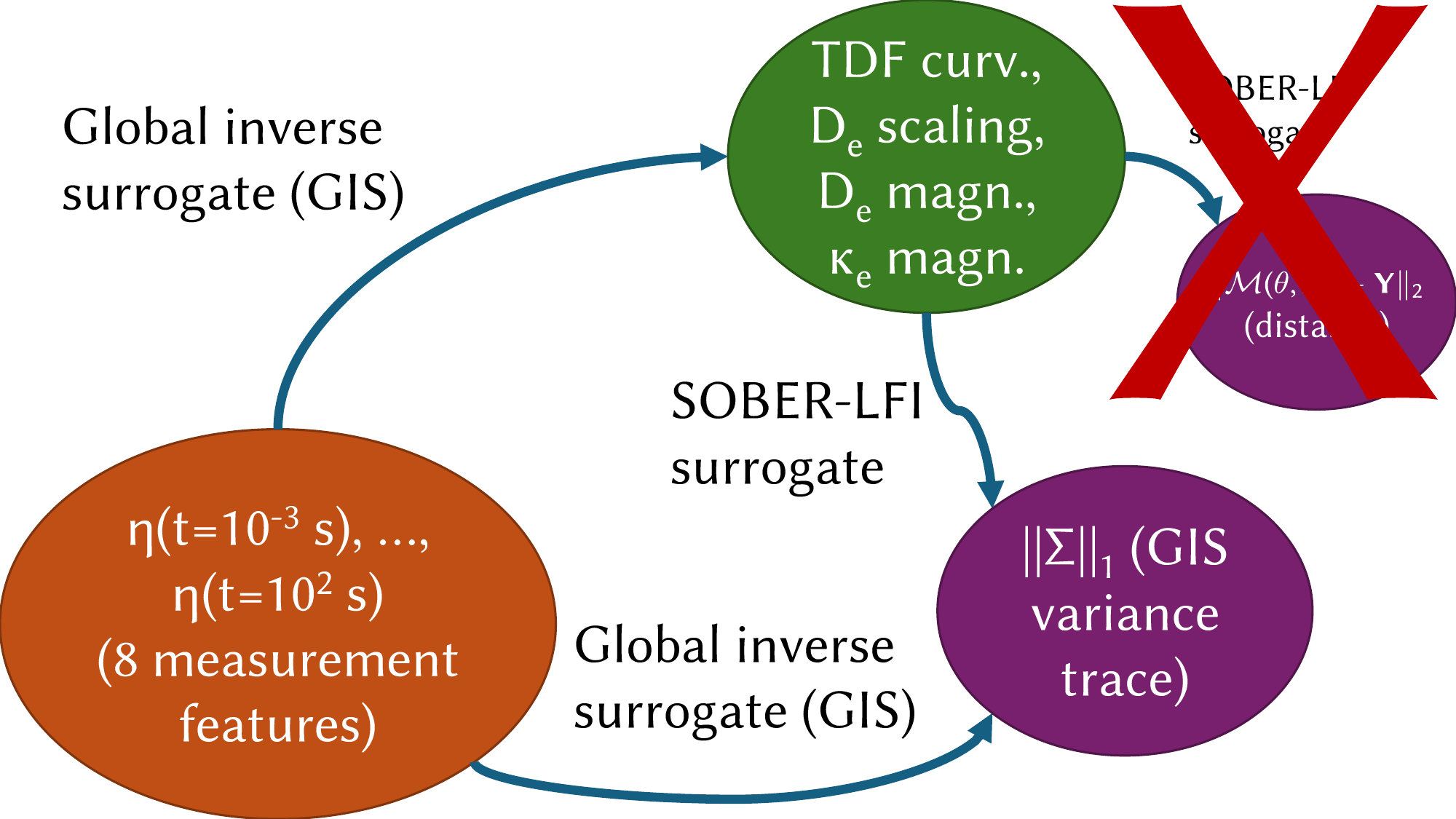}
    \caption{Illustration for the way SOBER is utilised within the global inverse surrogate training loop.}
    \label{fig:surrogate_encapsulation_high_dim}
\end{figure}

As our first example for advanced analysis, we apply our tools to assess reliability and accuracy of an electrochemical measurement technique. Our case study is the characterisation of electrolyte transport simply from measuring the electric potential under a constant current pulse. To analyse how well this simple approach works, we apply our global sensitivity analysis approach via a Gaussian process (GP) global inverse surrogate (GIS). We train this high-dimensional surrogate model to learn the relation between time-dependent overpotential under a constant current pulse and the electrolyte transport parameters. We use SOBER as an observer and guide to the training process, learning the relation between electrolyte transport parameters and difficulties in the training, see \prettyref{fig:surrogate_encapsulation_high_dim}. As an additional benefit, our approach accelerates parameterisation and allows further understanding of the characterisation process.

We wish to analyse whether or not a concentration dependence of the electrolyte diffusivity is relevant for these operating conditions, and if the electrolyte parameters are uniquely identifiable. Compared to the previous example 4, the global inverse surrogate here will have a considerably higher dimensionality, to show how to handle that in practice. As a basis, we recommend the reader reviews the previous examples on covariance and surrogate modelling.

This example considers the case where we have not yet acquired sufficient experimental data, and would like our physical model to inform us of any additional experiments we might need to perform. To emulate this without a laboratory at hand, this example contains no experimental data. Instead, we run forward simulations using the DFN model to generate synthetic overpotential data (with known electrolyte parameters).

With this ability to generate examples that we know the exact solution for, we now train a global inverse surrogate GP to `learn' the parameters from the synthetic overpotential-parameter data pairs. Importantly, we also use active learning (via SOBER) to decide which subsequent synthetic forward model runs to generate---this is to ensure we probe the most sensitive parts of the parameter space efficiently. The active learning approach has a highly beneficial side effect: the training data itself becomes structured according to the complexity of the inverse modelling task. Higher densities of training data indicate a challenging area of parameter space, and we can decide via our acquisition criterion which challenge the training data should tackle.
%
Once the global inverse surrogate GP is sufficiently trained, we can rapidly evaluate the sensitivity of any point in the parameter space via the covariance the GP outputs alongside the parameter prediction.

To detail the electrochemical experiment in our case study, we frame it with the following question: how much information can be gleaned about the electrolyte of a full cell from a single constant-current pulse? Especially considering the fact that electrolyte parameters are dependent on the electrolyte concentration \cite{Valoen2005}, does the dependence have a significant impact on the cell response? Sensitivity analysis (on the measurement) could answer this question for any one set of electrolyte parameters, but it could not tell us whether its results can be generalised to the whole range of physically sensible electrolyte parameter sets. This is where we instead train a global inverse surrogate to learn the information contained within an assortment of examples.

The electrolyte parameters that the global inverse surrogate model will have to predict are the thermodynamic factor TDF of the electrolyte, its diffusivity \(D_e\), and its conductivity \(\kappa_e\), according to the following usually observed relations \cite{Valoen2005} to the electrolyte concentration \(c_e\):
\begin{align}\begin{split}
    \text{TDF}(c_e) &= 1 + \text{TDF}_\text{curv.} \left(\frac{c_e}{c_{e,\text{reference}}}\right)^2, \\
    D_e(c_e) &= D_{e,\text{ magn.}} \exp\left(-D_{e,\text{ scaling}} \left(\frac{c_e}{c_{e,\text{ref.}}} - 1\right)\right), \\
    \kappa_e(c_e) &= \frac{\kappa_{e,\text{ magn.}}}{c_e/c_{e,\text{ref.}}} \exp\left(-\left(\frac{\log\left(\frac{c_e/c_{e,\text{ref.}}}{\kappa_e\text{ peak}}\right)}{\kappa_{e,\text{ spread}}}\right)^2\right).
\end{split}\end{align}

We fix \(\kappa_{e,\text{ peak}}\) and \(\kappa_{e,\text{ spread}}\) at 1.0, as we found them to be virtually unidentifiable in a first test run. Looking at a plot of \(\kappa_e(c_e)\), e.g. in Val{\o}en et al.\ \cite{Valoen2005} in Figure 14, it makes sense; any significant change to \(\kappa_e(c_e)\) for reasonable values of \(\kappa_{e,\text{ peak}}\) and \(\kappa_{e,\text{ spread}}\) happens far beyond sensible electrolyte concentration gradients. The range of the other four parameters we explore are defined in \prettyref{tab:electrolyte_identifiability}.
\begin{table}
\centering
\caption{Priors for electrolyte identifiability example; each is a multivariate log-uniform distribution defined as exactly confined within these bounds.}
\label{tab:electrolyte_identifiability}
\begin{center}
\begin{tabular}{c|ccc}
\hline
parameter & transform & bounds \\
\hline
\(\text{TDF}_\text{curv.}\) & log & \([0.1, 4.0]\) \\
\(D_{e,\text{ scaling}}\) & log & \([0.2, 2.0]\) \\
\(D_{e,\text{ magn.}}\) [\si{\square\metre\per\second}] & log & \([3\cdot10^{-10}, 10^{-9}]\) \\
\(\kappa_{e,\text{ magn.}}\) [\si{\siemens\per\metre}] & log & \([0.5, 2.0]\) \\
\hline
\end{tabular}
\end{center}
\end{table}

The input that the surrogate model will get for predictions is the overpotentials at eight logarithmically spaced timepoints between \qty{1}{\milli\second} and \qty{100}{\second} of a 5C constant-current pulse, i.e., a current 5 times higher than the battery capacity per hour. The overpotential evaluations are the difference between open-circuit voltage of the electrode materials and measured voltage of the battery. We again use the example battery parameters from the legacy DUALFOIL code \cite{Newman2007}. Note that these parameters are of a full graphite-cobalt oxide cell, so the overpotential includes the electrode-related voltage drops (e.g.\ caused by surface vs.\ average solid concentration, and kinetics).

In order to understand the structure of the resulting training data, we identified parameter groups of interest by K-Means clustering. We then performed sensitivity analysis on a representative parameter set from each group. We took 16 initial model samples and then ran 15 SOBER iterations at 32 samples each. The acquisition criterion was maximum uncertainty.

\paragraph{Result and Discussion}

\begin{figure}
    \centering
    \includegraphics[width=\columnwidth]{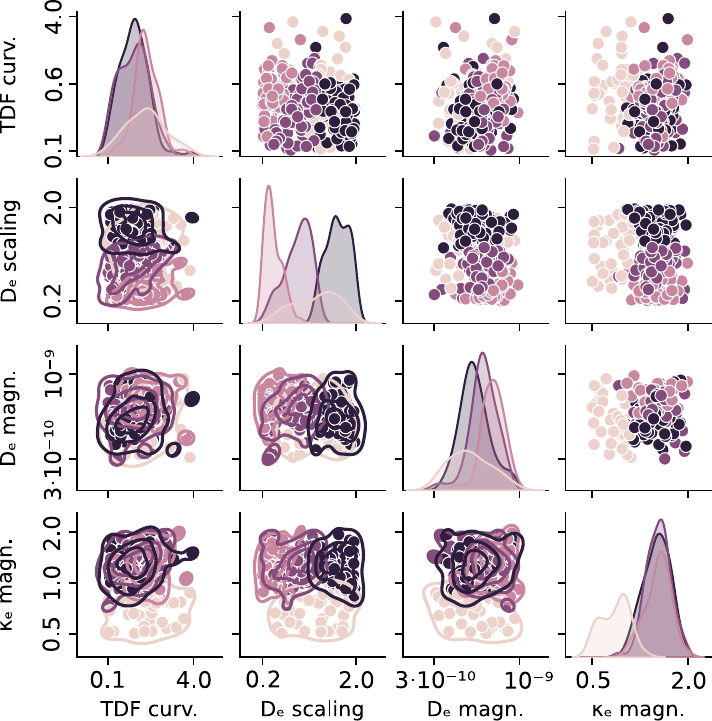}
    \caption{Training data---model parameters for inverse modelling example of electrolyte parameter identification. Groups of related parameter sets categorised via K-Means into 4 clusters, indicated by different colours.}
    \label{fig:diffusivity_training_parameters}
\end{figure}

We plot the resulting training dataset for our surrogate model GP in \prettyref{fig:diffusivity_training_parameters}, or rather, the parameters part of the (parameters, overpotentials) tuples. Overall, we evaluated around 500 different parameter combinations. We now wish to study the general accuracy of the electrolyte characterisation from rate tests. As our inverse model GP maps eight evaluations of the overpotential curve to four electrolyte parameters, the resulting \(320 = 8 [2 (4 + 16)]\)-dimensional object cannot be visualised directly: for each of the \(8\) outputs, we have \(4\) predicted inputs and their \(4 \times 4\) covariance matrices, plus \(4\) predicted sensitivities and their \(4 \times 4\) covariance matrices. 

\begin{figure}
    \centering
    \includegraphics[width=0.7\columnwidth]{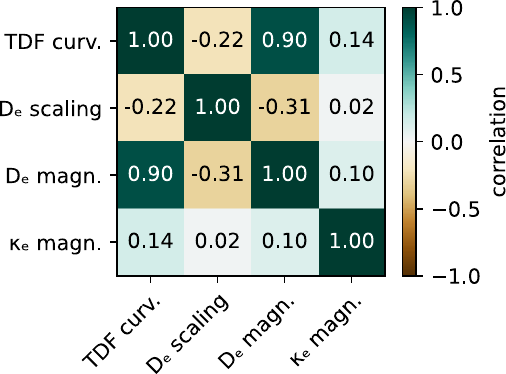}
    \caption{Correlation matrix of one representative evaluation of a Gaussian Process trained on \prettyref{fig:diffusivity_training_parameters} parameters. The strong cross-correlation between thermodynamic factor and diffusivity shows that they are interchangeable.}
    \label{fig:diffusivity_correlation_everywhere_else}
\end{figure}

Instead, we use the clustering of the model parameters part of the training data in \prettyref{fig:diffusivity_training_parameters}, reducing our high-dimensional parameter space to a limited number of clusters to study. Each cluster represents a region of parameter set space with an especially high model sensitivity. One way to think of each cluster centre is, perhaps, as representing a different electrolyte mixture. Instead of real electrolyte mixtures, we simulate the measured overpotential at the parameter values at the centres of each of the four clusters, and feed those simulations to the GP surrogate model. We can study global trends in the four corresponding correlation matrices. In this case, they follow the same trend every time, so we depict only the one of the most accurate prediction in \prettyref{fig:diffusivity_correlation_everywhere_else}. The number of clusters we chose in \prettyref{fig:diffusivity_training_parameters} may be determined by a metric like the silhouette score. But to understand the structure of the training dataset, it can be more helpful to look at several clusterings with different numbers of clusters and choose the number beyond which no meaningful new category arises.

To answer our first question about the diffusivity concentration dependence, we have to look at the covariance the surrogate reports for the test cases. Converted to two standard deviations credibility intervals, they turn out to cover the range of the prior of the training data, i.e., the concentration dependence of the electrolyte has no effect for the experimental protocol we used. To answer our second question about electrolyte identifiability, as we see in \prettyref{fig:diffusivity_correlation_everywhere_else}, there is a consistent cross-correlation between the thermodynamic factor and electrolyte diffusivity, showing us the importance of extracting one of those from another separate experimental measurement where they are identifiable separately from each other. But, tested on the parameters of the original electrolyte of the example cell, the surrogate reports perfect identifiability, with unity correlation matrices and small credibility intervals. Almost as a side result, the surrogate model GP we trained is now an extremely fast alternative way to process rate test data.

In contrast to the majority of machine learning applications, the presented active learning approach minimises the amount of training data required. Therefore, this approach may be applied to real data in automated laboratories. This way, we would efficiently validate a given experimental procedure not on singular test cases, but on a global view obtained from the most challenging experiments.

\subsection{Example 6: Impedance model selection}

\paragraph{Method}

Data-informed model development and selection is another advanced technique that our methodology supports. We will consider various electrochemical impedance spectroscopy (EIS~\cite{Talian2024}) models for two reasons: to prove the validity of our approach on a currently relevant area of research and to showcase the robustness of our methodology in the face of challenging experimental conditions. %
%
This sixth and final example showcase focuses on the concept of evidence, using covariance only as a means of providing error bars. We'll learn how to decide on a specific cell model based on evidence on impedance data, and how to interpret said evidence if the given data barely supports a confident decision.

The work of Hallemans et al.\ \cite{Hallemans2024} allows us to use DFN impedance simulations for parameterisation, and said work verifies that it works perfectly and directly on synthetic data via fitting the Nyquist plots. Our complement generalises this work to real battery data, as noise in the data and remaining uncertainty in various model parameters hinder the direct curve fitting approach. For context, you may be aware that EIS data are most commonly fitted using equivalent circuit models (ECM). Our model selection method for ECMs (and the related transmission line models) was already given in an earlier publication \cite{Adachi2023a}. In an impedance Nyquist plot, building an ECM is building a shape out of semicircles and lines. But if we just reproduce the geometrical shapes for prediction purposes, we lack the physical connection to material properties. More involved ECMs are motivated by the physical phenomena inside a battery, but then they still only approximate a DFN impedance spectrum and may distort parameter interactions. But with the work of Hallemans et al.\ \cite{Hallemans2024}, we can simulate DFN impedance spectra via automatic differentiation. This approach allows us to use our inverse modelling approach with the DFN instead of ECMs, as the simulations are stable and quick enough for a sensible run-time. Therefore, we will not consider ECMs going forward. We combine the DFN impedance model with a compatible thermodynamically consistent impedance model for the solid electrolyte interphase (SEI) \cite{Single2019}, as simulating SEI impedance within the DFN is not yet feasible. %
%
The first of the two EIS datasets we will consider covers the analysis of mesostructure effects. We use impedance data from an NMC electrode we gratefully received from Günther et al.\ \cite{Günther2025}, see \prettyref{fig:gunther}. We extend their qualitative analysis of the influence of electrode mesostructure with our quantitative approach.

The parameter priors are set up per \prettyref{tab:mercedes_impedance}. They correspond to the most prominent features in the impedance: diffusivities in active material \(D_\text{s}\) and electrolyte \(D_\text{e}\), exchange-current density \(i_0\) and double-layer capacitance \(C_\text{DL}\),  Bruggeman coefficient \(\beta\) of the electrode, and electrode electronic conductivity \(\sigma_\text{s}\). Note that we copy the observation from Günther et al.\ \cite{Günther2025} that the thicker electrodes partly pulverised during preparation, which we reflect with a rather low electronic conductivity due to defects in the binder network. %
\begin{table}
\centering
\caption{Priors for electrolyte impedance model example---a multivariate mixture of uniform and log-uniform priors defined as exactly confined within these bounds. The SPM parameterisations consider the first three parameters, and the SPMe and DFN parameterisations consider all six.}
\label{tab:mercedes_impedance}
\begin{center}
\begin{tabular}{c|ccc}
\hline
parameter & transform & bounds \\
\hline
\(D_\text{s}\) [\si{\metre\per\square\second}] & log & \([10^{-17}, 3\cdot10^{-16}]\) \\
\(C_\text{DL}\) [\si{\farad\per\square\metre}] & log & \([0.1, 0.5]\) \\
\(i_0\) [\si{\ampere\per\square\metre}] & log & \([0.01, 0.03]\) \\
\arrayrulecolor{gray}\hline
\(\beta\) & -- & \([3.5, 5.0]\) \\
\(\sigma_\text{s}\) [\si{\siemens\per\metre}] & log & \([0.2, 1.0]\) \\
\(D_\text{e}\) [\si{\metre\per\square\second}] & log & \([10^{-10},  10^{-9}]\) \\
\arrayrulecolor{black}\hline
\end{tabular}
\end{center}
\end{table}
%
In accordance with the work of Günther et al.\ \cite{Günther2025}, we expect the SPM and the SPMe or the DFN to produce similar fits on the data for which the SPM suffices, but with lower evidence for the superfluous SPMe or DFN. Additionally, with increasing electrode thickness, this assessment should flip since only the DFN can capture some mesostructure effects, with the SPMe being somewhere in between.

The second of the two datasets we consider focuses on timescales typically attributed to SEI and double-layer effects. We use impedance data on a graphite electrode from one of our earlier publications \cite{Kuhn2025a}. We investigate whether the timescales for SEI and electrochemical double-layer can be uniquely assigned from the data alone. To do so, we use the same data and the same DFN+SEI model twice and instead vary the priors as per \prettyref{tab:SEI_impedance}. One prior establishes the timescale of the electrochemical double-layer as faster than that of the SEI, and the other reverses that order. We vary four parameters that correspond to those timescales, which for the double-layer are its capacitance \(C_\text{DL}\) and the exchange-current density \(i_0\), and for the SEI they are its relative permittivity \(\varepsilon_\text{SEI}\) and its Bruggeman coefficient \(\beta_\text{SEI}\). %
\begin{table}
\centering
\caption{Priors for SEI and double-layer impedance model example. Top four rows correspond to a double-layer timescale faster than the SEI, and vice versa for bottom four rows. Priors are a multivariate mixture of normal and log-normal, defined such that 95\% confidence regions touch these bounds.}
\label{tab:SEI_impedance}
\begin{center}
\begin{tabular}{c|ccc}
\hline
parameter & transform & bounds \\
\hline
\(C_\text{DL}\) [\si{\farad\per\square\metre}] & log & \(7\cdot[10^{-4}, 10^{-3}]\) \\
\(i_0\) [\si{\ampere\per\square\metre}] & log & \([0.02, 0.08]\) \\
\(\varepsilon_\text{SEI}\) & log & \([100, 200]\) \\
\(\beta_\text{SEI}\) & -- & \([4.3, 4.8]\) \\
\arrayrulecolor{gray}\hline
\(C_\text{DL}\) [\si{\farad\per\square\metre}] & log & \([0.01, 0.1]\) \\
\(i_0\) [\si{\ampere\per\square\metre}] & log & \([0.02, 0.08]\) \\
\(\varepsilon_\text{SEI}\) & log & \([10, 50]\) \\
\(\beta_\text{SEI}\) & -- & \([4.3, 4.8]\) \\
\arrayrulecolor{black}\hline
\end{tabular}
\end{center}
\end{table}
%
We expect the SEI model to score a higher model evidence when placed at the lower-frequency end, consistent with literature results \cite{Horstmann2019}.

To preprocess impedance data for parameterisation, distribution of relaxation times (DRT) \cite{IversTiffee2017} is the most general tool to filter noise and reduce data to distinct features. The DRT features are peaks in the DRT frequency spectra. See Zhao et al.\ \cite{Zhao2022} for an application of DRT to diffusion phenomena. As for the cost function, we define the distance between two DRT spectra as the sum over the distances of each peak to its nearest peak in the other spectrum in a double-logarithmic plot. For all impedance examples, regardless of number of fit parameters, we take 48 initial model samples and then run 19 SOBER iterations at 48 samples each. We specify the acquisition function as the batch version of negative integrated posterior variance to minimise the evidence calculation error. The BASQ posterior evaluation is set up each time with 3 to the power of the number of fit parameters many integration nodes.

\paragraph{Result and Discussion}

\begin{figure}
    \begin{center}
        \includegraphics[width=\columnwidth]{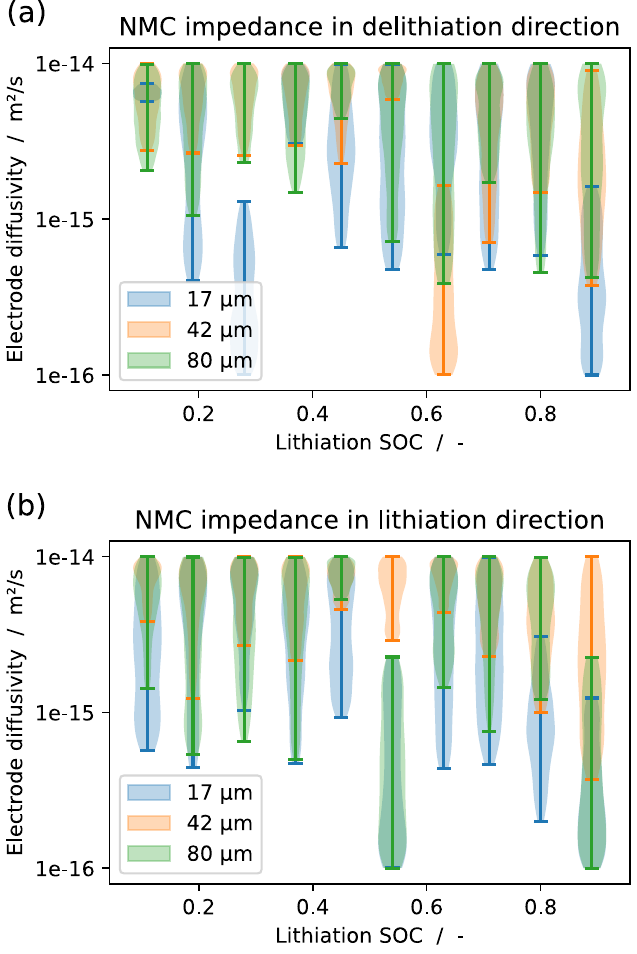}
    \end{center}
    \caption{Violin plots of active material diffusivities as fitted with DFN model on electrodes with varying thickness: (a) delithiation and (b) lithiation before impedance measurement. Widths in SOC-direction indicate probability density at that diffusivity value. The vertical error bars are the one-sigma credibility intervals, containing 68\% of the probability distributions. Convergence is verified via predictive posterior plots such as in \prettyref{fig:impedance_80_microns}.}
    \label{fig:impedance_full}
\end{figure}

For the mesostructure case study, we show the fitted results for the active material diffusivity with the DFN model in \prettyref{fig:impedance_full}. These serve for comparison with Figure 21 in the original study of Günther et al.\ \cite{Günther2025}, where we see a disparity of an order of magnitude in measured diffusivities between the \qty{80}{\micro\metre} electrode and the other two electrodes. To clarify, there the underlying GITT data was used as such, evaluated with the method of Kang et al.\ \cite{Kang2021}. Günther et al. also performed a Fourier transform of the GITT data, producing the impedance spectra that we use here, but they didn't use them for diffusivity characterisation.

\begin{figure}
    \begin{center}
        \includegraphics[width=\columnwidth]{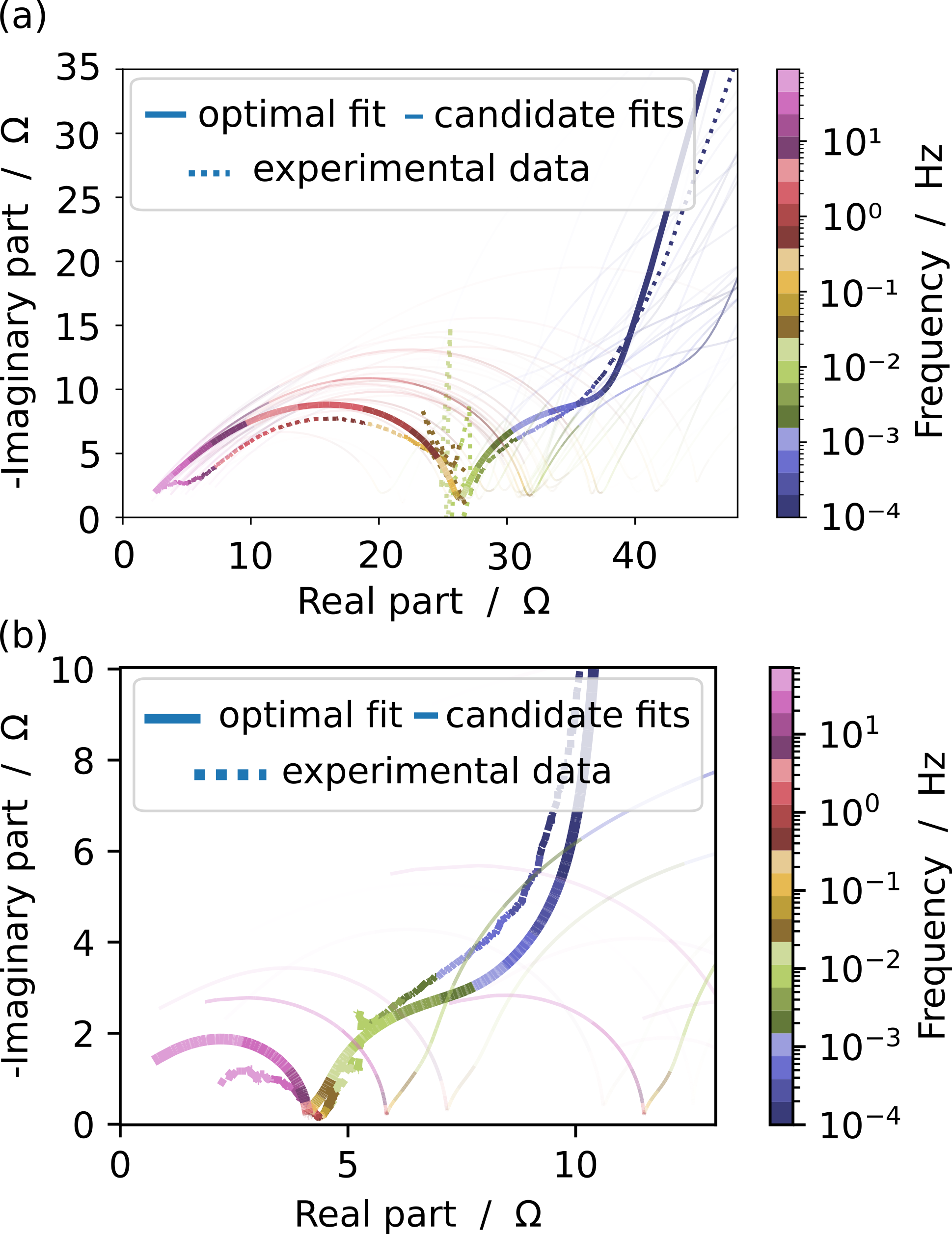}
    \end{center}
    \caption{Predictive posterior of impedance parameterisation with DFN model on \qty{80}{\micro\metre} electrode at (a) 89\% lithiation (in delithiation direction) and (b) 45\% lithiation (in lithiation direction). Since the purely resistive offset is not part of the DRT spectra and was hence fitted poorly, we adjusted the offset of the simulations in plot (b) for increased clarity.}
    \label{fig:impedance_80_microns}
\end{figure}

Here, we combined our DRT spectra distance approach, fast and stable DFN impedance simulations, and SOBER to characterise diffusivities from the impedance spectra. With this, we can give an explanation to the diffusivity disparity. In the original study~\cite{Günther2025} for the data, no error bars were given. But with the credibility intervals in \prettyref{fig:impedance_full} that we can calculate via SOBER, we now see that the measurement simply does not allow us a better characterisation of the diffusivity than within one order of magnitude. Of course, we need to verify that our parameterisation converged to make this claim. We validate convergence via the predictive posteriors, of which we plot two exemplary ones in \prettyref{fig:impedance_80_microns}. The narrow band of good fits that also aligns with what SOBER predicts to be a good fit confirms convergence.

These two measurement points reflect different challenges in the measurement procedure, as reported by Günther et al.\ \cite{Günther2025}. In the preparation of the two thicker electrodes, parts of the electrodes were pulverised, which affected the measurements at high frequency, see \prettyref{fig:impedance_80_microns}(a) at 89\% lithiation. Additionally, the characteristic frequency of the measurement setup may introduce noise at exactly those frequencies at which SPM and DFN diverge, see \prettyref{fig:impedance_80_microns}(b) at 45\% lithiation. Still, measurement accuracy is generally better away from the SOC extremes. We now analyse how those challenges may be detected automatically via our uncertainty quantification approach.

We calculate the evidences for both measurement points in \prettyref{tab:impedance_evidence}. As the evidences suffer from large variances, we show one standard deviation credibility intervals. For a complete picture, we plot the evidence distributions in \prettyref{fig:impedance_evidences}. As long as we ensure the variances are smaller than the means, we may still rank the models by their mean evidences \cite{Adachi2023a}.
\begin{table*}
\centering
\caption{One standard deviation credibility intervals (square brackets) of evidences and relative comparisons between mean evidences (round brackets)  computed for impedance models. Impedance models include (DFN) or ignore (SPM) electrolyte concentration gradients and are fitted on measurements with varying positive electrode thickness. For display purposes, all evidences are multiplied by 1000.}
\label{tab:impedance_evidence}
\vspace{11pt}
\begin{supertabular}{cgcgc}
\begin{minipage}[c]{0.2\columnwidth}
    \begin{center}
        \phantom{\textbf{quality}}
    \end{center}
\end{minipage}
&
\begin{minipage}[c]{0.8\columnwidth}
    \begin{center}
        \textbf{89\% lithiation\\(during delithiation)}
        \vspace{11pt}
    \end{center}
\end{minipage}
&
\begin{minipage}[c]{0.8\columnwidth}
    \begin{center}
        \textbf{45\% lithiation\\(during lithiation)}
        \vspace{11pt}
    \end{center}
\end{minipage} \\

\arrayrulecolor{lightgray}\hline\arrayrulecolor{.}

\begin{minipage}[c]{0.2\columnwidth}
    \begin{center}
        \textbf{quality\\of\\data}
    \end{center}
\end{minipage}
&
\cellcolor{lightgray}\begin{minipage}[c]{0.8\columnwidth}
    \vspace{11pt}\begin{center}
        \begin{tabular}{c|ccc}
            \hline
            (a)\phantom{aaa} & \qty{17}{\micro\metre} & \qty{42}{\micro\metre} & \qty{80}{\micro\metre} \\
            \hline
            SPM & \([0.6, 95]\) & \([2.1, 16]\) & \([0.5, 35]\) \\
             & (100\%) & (76\%) & (55\%) \\
            DFN & \([0.02, 2.8]\) & \([0.3, 0.6]\) & \([0.3, 17]\) \\
             & (100\%) & (175\%) & (863\%) \\
            \hline
        \end{tabular}
    \end{center}\vspace{0pt}
\end{minipage}
&
\begin{minipage}[c]{0.8\columnwidth}
    \vspace{11pt}\begin{center}
        \begin{tabular}{c|ccc}
            \hline
            (b)\phantom{aaa} & \qty{17}{\micro\metre} & \qty{42}{\micro\metre} & \qty{80}{\micro\metre} \\
            \hline
            SPM & \([224, 1085]\) & \([0.8, 61]\) & \([3, 138]\) \\
             & (100\%) & (1.4\%) & (4.3\%) \\
            DFN & \([207, 217]\) & \([0.01, 7]\) & \([0.1, 11]\) \\
             & (100\%) & (0.1\%) & (0.6\%) \\
            \hline
        \end{tabular}
    \end{center}\vspace{0pt}
\end{minipage} \\

\arrayrulecolor{lightgray}\hline\arrayrulecolor{.}

\begin{minipage}[c]{0.2\columnwidth}
    \begin{center}
        \textbf{quality\\of\\model}
    \end{center}
\end{minipage}
&
\begin{minipage}[c]{0.8\columnwidth}
    \vspace{11pt}\begin{center}
        \begin{tabular}{c|cc}
            \hline
            (c)\phantom{aaa} & SPM & DFN (corr.) \\
            \hline
            \qty{17}{\micro\metre} & \([0.6, 95]\) & \([0.3, 45]\) \\
             & (100\%) & (51\%) \\
            \qty{42}{\micro\metre} & \([2.1, 16]\) & \([5, 9]\) \\
             & (76\%) & (117\%) \\
            \qty{80}{\micro\metre} & \([0.5, 35]\) & \([4, 265]\) \\
             & (55\%) & (792\%) \\
            \hline
        \end{tabular}
    \end{center}\vspace{0pt}
\end{minipage}
&
\cellcolor{lightgray}\begin{minipage}[c]{0.8\columnwidth}
    \vspace{11pt}\begin{center}
        \begin{tabular}{c|ccc}
            \hline
            (d)\phantom{aaa} & SPM & DFN (corr.) \\
            \hline
            \qty{17}{\micro\metre} & \([224, 1085]\) & \([3256, 3424]\) \\
             & (100\%) & (676\%) \\
            \qty{42}{\micro\metre} & \([0.8, 61]\) & \([0.2, 107]\) \\
             & (1.4\%) & (49\%) \\
            \qty{80}{\micro\metre} & \([3, 138]\) & \([2, 177]\) \\
             & (4.3\%) & (94\%) \\
            \hline
        \end{tabular}
    \end{center}\vspace{0pt}
\end{minipage} \\
\end{supertabular}
\end{table*}

We have two ways to compare the mean evidences. The first can only compare each model with itself on varying data. Therefore the percentages in \prettyref{tab:impedance_evidence} for data quality relate the mean evidences to the \qty{17}{\micro\metre} data, which we chose since SPM and SPMe or DFN produce virtually the same result when no mesostructure effects are relevant. The second way compares different models on the same data. To do so, we have to correct for the integral penalty for the extra 3 fit parameters for the DFN, multiplying the evidence by \(\sqrt{2 \pi}^3\). The factor \(\sqrt{2 \pi}\) stems from the normalisation factor of a (multivariate) Gaussian distribution \(\mathcal{N}(\mu, \Sigma)\) with dimensionality \(d\): \(\sqrt{(2 \pi)^d \det(\Sigma)}^{-1} \exp(-0.5(x - \mu)^T \Sigma^{-1}(x - \mu))\). Identifiable optimisation tasks tend to a Gaussian posterior asymptotically per the central limit theorem. We prefer comparing the determinant of \(\Sigma\) between different models directly, without the \textquote{arbitrary} dimensionality penalty \(\sqrt{2 \pi}^{d_2 - d_1}\). The percentages in \prettyref{tab:impedance_evidence} for model quality directly relate those evidences to their SPM counterparts, each for the same electrode thickness.

The computed evidences track the experimental observations in the original paper \cite{Günther2025} for the data. In \prettyref{tab:impedance_evidence}(a), we see that the electrode-only single particle model decreases in confidence with increasing electrode thickness, which introduces mesostructure effects only captured in the DFN. \prettyref{tab:impedance_evidence}(c) shows that for \qty{17}{\micro\metre}, the DFN is indeed superfluous, while it is more appropriate than the SPM for the thicker electrodes. As for \prettyref{tab:impedance_evidence}(b), we see a different picture. The evidences are generally higher than in \prettyref{tab:impedance_evidence}(a), which is in line with better measurement accuracy away from the SOC extremes. More surprising is that the \qty{17}{\micro\metre} data has much higher evidence than the two thicker electrodes for both SPM and DFN. Still, the electrolyte part of the parameterisation works, albeit it is penalised for not being very stable. And with the calibration in \prettyref{tab:impedance_evidence}(d), we recover the expected model choices for the thicker electrodes.

We may formally interpret the results from \prettyref{tab:impedance_evidence} by considering what these evidences precisely mean. Applying Bayes' rule on the model evidence \(p(\mathbf{D} \mid \mathcal{M})\) allows us to factorise it:
\begin{equation}
    p(\mathbf{D} \mid \mathcal{M}) = \frac{p(\mathcal{M} \mid \mathbf{D}) p(\mathbf{D})}{p(\mathcal{M})}. \label{eq:evidence}
\end{equation}

Since in \prettyref{tab:impedance_evidence}(a,c), the dominating measurement difficulty was the extreme SOC \cite{Günther2025}, the quality of the data \(p(\mathbf{D})\) was similar across electrodes. Without prior bias towards any one model, we may assign equal probabilities \(p(\mathcal{M})\). Hence both \(p(\mathbf{D} \mid \mathcal{M})\) and \(p(\mathcal{M} \mid \mathbf{D})\) have similar numerical values, which is why we can use the evidence as a measure of the model quality given the data. In \prettyref{tab:impedance_evidence}(b,d), on the other hand, the issues in preparing the two thicker electrodes as pristinely as the thinnest one dominate the evidence, so we have to interpret it as the measure of data quality given the model.

\begin{figure}[!ht]
    \begin{center}
        \includegraphics[width=\columnwidth]{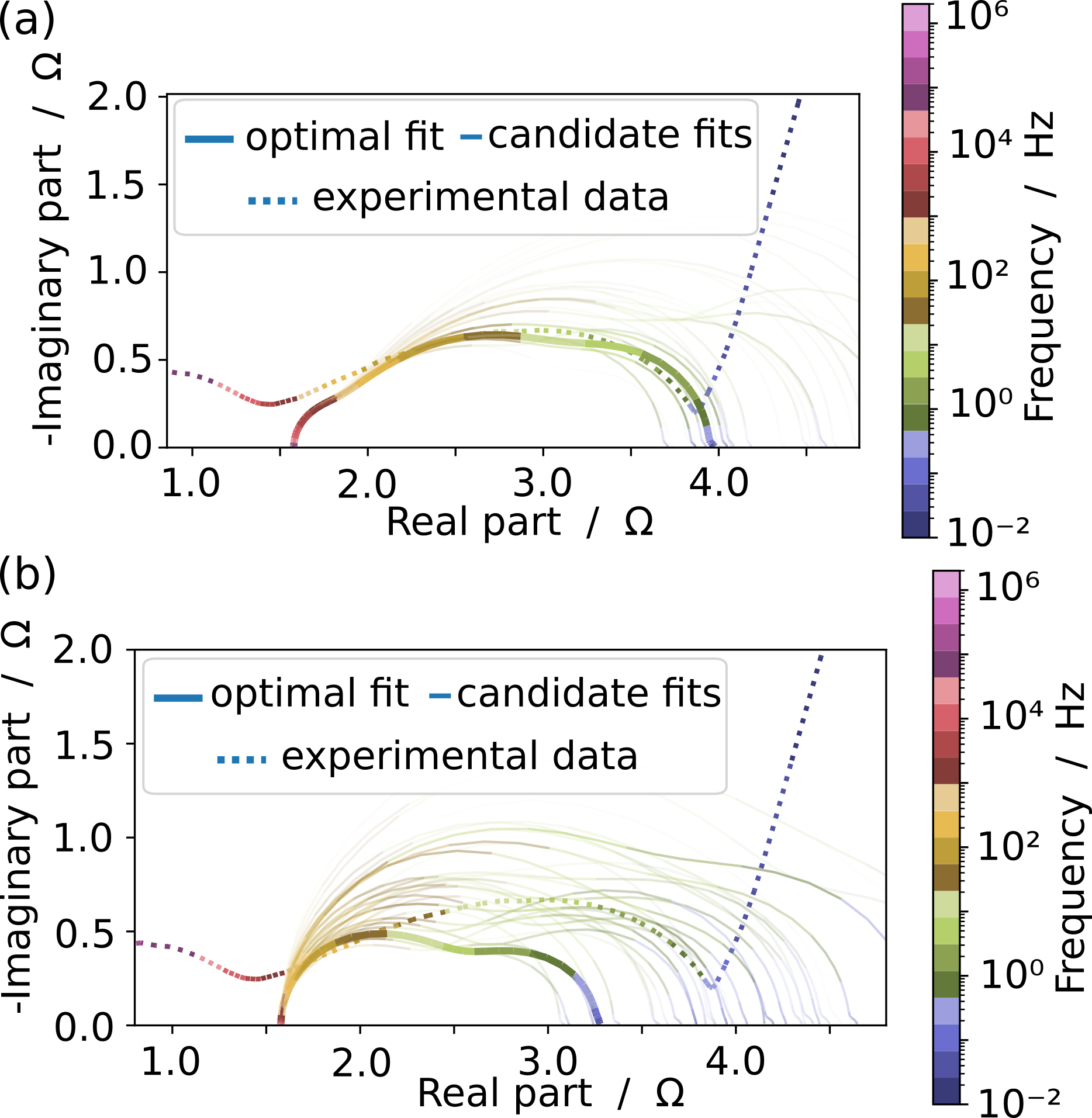}
    \end{center}
    \caption{Predictive posterior of impedance parameterisation on negative electrode of LG MJ1 18650 cell with (a) a prior bias towards double-layer timescale faster than SEI and (b) a prior bias towards double-layer timescale slower than SEI. The high-frequency semicircle at about \qty{e5}{\hertz} and the low-frequency diffusivity tail are intentionally cut out of the DRT spectra and not part of the model.}
    \label{fig:lg_mj1_predictive_posterior}
\end{figure}

With our impedance model analysis method verified, we can repeat it on the other example dataset that more closely looks at the timescales attributed to the SEI. We show the respective predictive posteriors in \prettyref{fig:lg_mj1_predictive_posterior}. The mean evidences with one standard deviation are \([0.04, 0.17]\) and \([6\cdot10^{-6}, 2\cdot10^{-3}]\) (see the complete underlying distributions in \prettyref{fig:impedance_evidences}) for the prior with a bias towards a faster double-layer timescale than that of the SEI and vice versa, respectively.

The evidences clearly assign the higher frequencies to the double-layer and the lower frequencies to the SEI. If one is familiar with the general shapes of the underlying models, they may have expected it to be the other way around. This shows how a complete fit of the interphase phenomena greatly benefits us compared to a manual analysis of the individual semicircles and Warburg shapes. The extreme frequencies relate to electrode conduction and diffusivity and are not modelled here to focus the evidence on the interphase \cite{Talian2024}.

In conclusion, we have established the efficacy and validity of our Bayesian model selection approach for real impedance data. With the model selection calculations, we automatically also get parameterisations with precise quantification of remaining uncertainties. And with the various predictive posteriors associated with the condensed evidences, we start to understand the link between evidence and the uncertainty in a model parameterisation.

\section{Conclusion}

Characterisation of long-term battery phenomena has been demonstrated with many machine learning approaches on existing chemistries. For new chemistries, a factorially growing number of arbitrarily deeply nested battery models has been created. Using SOBER and BASQ, Bayesian approaches for model parameterisation and selection, we can automatically, accurately, and quickly sort through this plethora of models.

SOBER can handle previously intractable parameterisation and model selection tasks on highly coupled or non-linear battery models. Additionally, BASQ can estimate the global suitability of a model by considering all feasible model realisations. Even if several models may be tuned to produce the same result, global parameter analysis via Bayesian model selection can discern which model does so most reliably and parsimonously. To support this claim, we showcased a representative selection of application examples on mechanical voltage relaxation, knee point identification, cycling data augmentation, electrolyte identifiability analysis, and impedance model selection.

To make SOBER more accessible to the broader battery community, we wrapped a concise interface to it within PyBOP~\cite{Planden2024}. Advanced techniques may be appended within the same framework as required.

Future usage of SOBER may help us find new competitive battery chemistries by accelerating their model-based study. The examples we use here are a representative showcase of the capabilities of SOBER, but far from its only applications. Embedding techniques can generalise SOBER to optimisations on high-dimensional spaces, as showcased in its original publication \cite{Adachi2024}. With support for massively parallelised computations, SOBER and BASQ can be utilised as a comprehensive search algorithm for models of novel battery compositions or materials.

\section*{Code availability}

\urlstyle{same}

The Bayesian Optimisation algorithm SOBER with the Bayesian Quadrature algorithm BASQ is available Open Source at \url{https://github.com/ma921/SOBER} or via Python package installation \textquote{pip install sober-bo}. The code to run the example applications is available Open Source within \url{https://github.com/pybop-team/PyBOP}, which you may install without the examples via \textquote{pip install pybop}.

\section*{Acknowledgements}

This work was supported by the German Aerospace Center (DLR). We thank Jonas Günther and Dominik Wycisk from Mercedes for providing and assisting with their experimental data. We thank Dennis Kopljar from DLR for inspiring the electrolyte identifiability showcase. The authors acknowledge support by the Helmholtz Association through grant no KW-BASF-6 (Initiative and Networking Fund as part of the funding measure \textquote{ZeDaBase-Batteriezelldatenbank}). This work contributes to the research performed at CELEST (Center for Electrochemical Energy Storage Ulm-Karlsruhe).

\section*{Author contributions}

\textbf{Yannick Kuhn}: Conceptualisation, Data curation, Formal analysis, Investigation, Methodology, Project administration, Resources, Software, Validation, Visualisation, Writing -- original draft, Writing -- review \& editing.
\textbf{Masaki Adachi}: Conceptualisation, Methodology, Resources, Software, Writing -- original draft, Writing -- review \& editing.
\textbf{Micha Philipp}: Writing -- review \& editing.
\textbf{David A. Howey}: Conceptualisation, Funding acquisition, Project administration, Supervision, Writing -- review \& editing.
\textbf{Birger Horstmann}: Conceptualisation, Funding acquisition, Project administration, Supervision, Writing -- review \& editing.

\section*{Conflicts of interest}

The authors declare no conflict of interest.

\section{ORCID}

\urlstyle{same}
\noindent
Yannick Kuhn \includegraphics[height=\baselineskip]{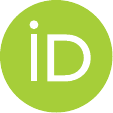} \url{https://orcid.org/0000-0002-9019-2290} \\
Masaki Adachi \includegraphics[height=\baselineskip]{Orcid_logo.pdf} \url{https://orcid.org/0000-0003-2580-2280} \\
David A. Howey \includegraphics[height=\baselineskip]{Orcid_logo.pdf} \url{https://orcid.org/0000-0002-0620-3955} \\
Birger Horstmann \includegraphics[height=\baselineskip]{Orcid_logo.pdf} \url{https://orcid.org/0000-0002-1500-0578} \\

\section*{Appendix}

\renewcommand{\thefigure}{A-\arabic{figure}}
\setcounter{figure}{0}

\paragraph{Utilising EP-LFI for preconditioning misspecified priors}

Expectation propagation (EP~\cite{Minka2001}) was originally proposed as a recursive version of moment-matching. Moment-matching assumes that the posterior distribution factorises into some set of tractable distributions, one per data point/feature. The posterior then results from a sequential pass over the data, similar to a Kalman filter. EP removes dependence on data order by recursively passing over it, similar to a smoothing Kalman filter. Critically, moment-matching and EP differ from Kalman filters in that they do not incorporate new data. Barthélme et al.\ \cite{Barthelme2014} built upon EP for likelihood-free inference (LFI) and made it into a form of importance sampling: EP-LFI. The benefit here is that, with LFI, you usually end up discarding a lot of samples in high-dimensional optimisation settings. This is due to the fact that a truly high-dimensional cost function will be far from its optimal value almost everywhere, reducing acceptable samples to a narrow region around a few points. But with the types of settings encountered in battery science, the issue is most often an overly wide prior. By introducing features tuned from expert knowledge, the cost function for each feature may have near-optimal values around a much higher-dimensional subset of the parameter space, allowing many more samples to inform the posterior calculation. Additionally, the interim posteriors constrict the search space for the following features to only those parameters that do not already contradict an earlier feature. So we want to employ EP-LFI mainly in the case of overly wide priors. For convergence, we may take the posterior of a still-suitably-wide EP-LFI run and use it as the prior for a non-EP run. The end result is asymptotically guaranteed to be the same, except for a multimodal true posterior. EP-LFI converges quadratically to unimodal posteriors, just as Newton's method, and in the same way, has to be dampened by only using a fraction of the full step/Bayesian update each time. A good rule of thumb is to adjust EP-LFI so that the final result will still contain half of the prior. Given a final dampening \(\alpha\) (e.g., 0.5) for \(K\) passes through all features, the dampening \(\alpha_f\) of each feature update has to be:
\begin{equation}
    \alpha_f = 1 - J (1 - \sqrt[JK]{\alpha}).
\end{equation}

\printbibliography

\begin{figure}
    \centering
    \includegraphics[width=\columnwidth]{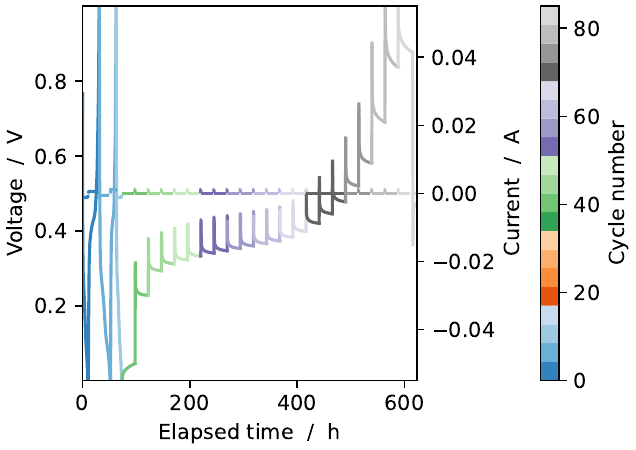}
    \caption{The GITT protocol from which we analyse one representative relaxation phase from Wycisk et al.\ \cite{Wycisk2024}.}
    \label{fig:wycisk}
\end{figure}
\begin{figure}
    \centering
    \includegraphics[width=\columnwidth]{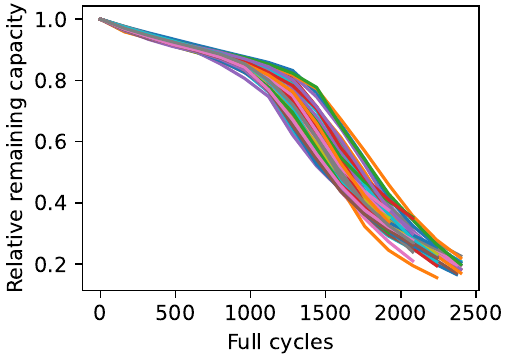}
    \caption{Battery degradation data on 48 identical cells from Baumhöfer et al.\ \cite{Baumhöfer2014}.}
    \label{fig:baumhofer}
\end{figure}
\begin{figure}
    \centering
    \includegraphics[width=0.98\columnwidth]{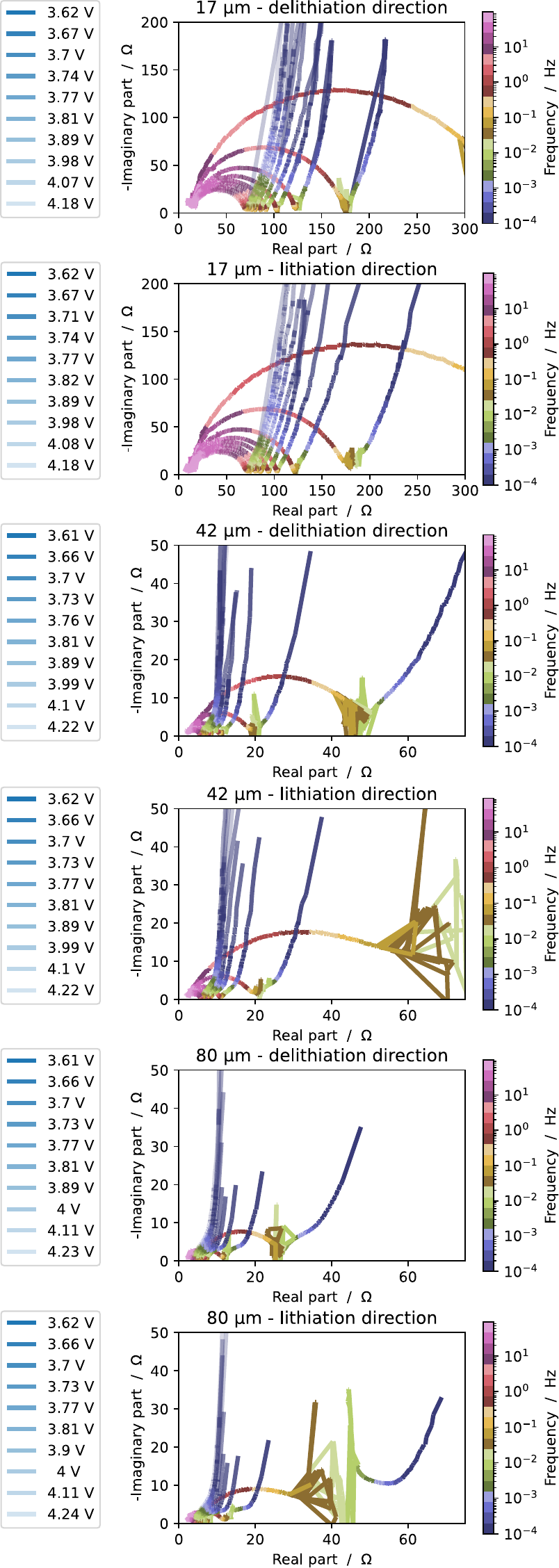}
    \caption{The impedance dataset which we analyse for NMC diffusivity signals from Günther et al.\ \cite{Günther2025} on NMC electrodes of varying thickness.}
    \label{fig:gunther}
\end{figure}
\begin{figure}
    \centering
    \includegraphics[width=\columnwidth]{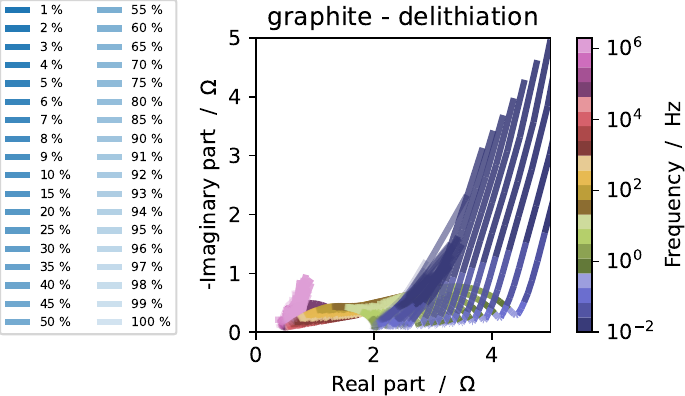}
    \caption{The impedance dataset which we analyse for SEI and double layer signals from Kuhn et al.\ \cite{Kuhn2025a}. Percentages describe lithiation between the minimal and maximal lithiation per manufacturer specifications.}
    \label{fig:kuhn}
\end{figure}
\begin{figure}
    \centering
    \includegraphics[width=0.8\columnwidth]{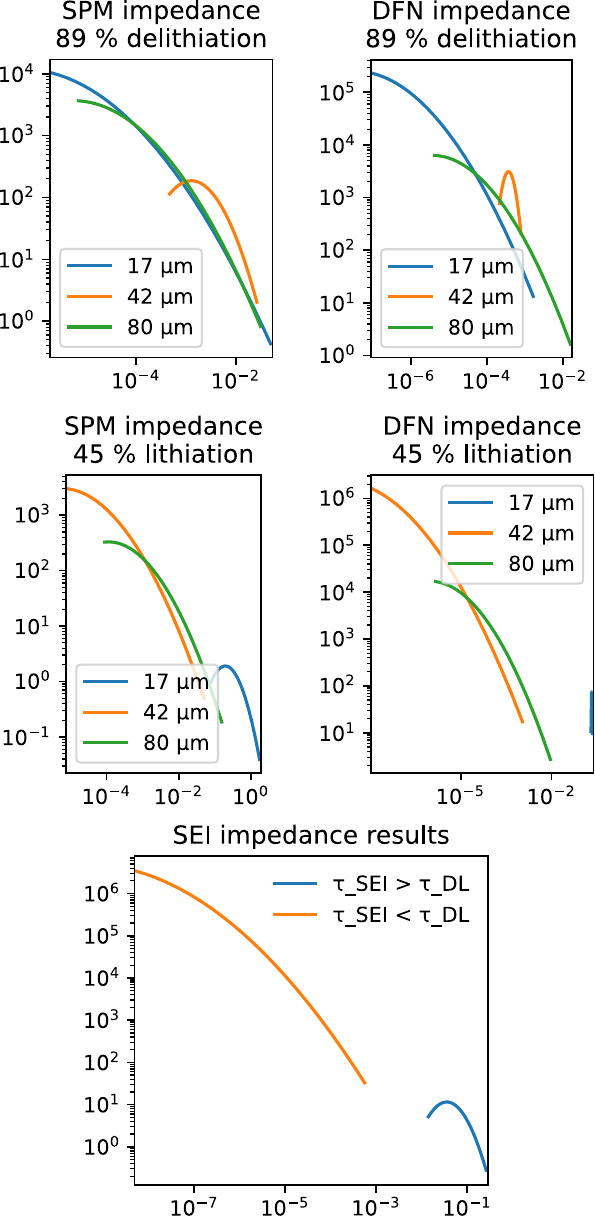}
    \caption{Evidence distributions of the impedance parameterisations of varying cell thicknesses with the SPM and DFN, as well as evidence distributions of the impedance parameterisations of an LG MJ1 18650 cell with the DFN+SEI model with two priors with different order between double-layer timescale and SEI timescale.}
    \label{fig:impedance_evidences}
\end{figure}

\end{document}